\documentclass[aps,twocolumn,floatfix,nofootinbib,superscriptaddress]{revtex4-1}
\usepackage{amssymb,amsmath}
\usepackage[mathscr]{euscript}
\usepackage{graphicx,cases}
\usepackage{epsfig}
\usepackage{textcomp}
\usepackage{epstopdf}
\usepackage{float}
\usepackage{braket}
\usepackage[normalem]{ulem}
\usepackage{enumerate}
\usepackage{graphicx}
\usepackage{psfrag}
\usepackage{enumerate}
\usepackage[makeroom]{cancel}
\usepackage{xcolor}

\usepackage[%
  colorlinks=true,
  urlcolor=blue,
  linkcolor=blue,
  citecolor=blue
]{hyperref}

\def\beq{\begin{equation}}
\def\eeq{\end{equation}}
\def\bea{\begin{eqnarray}}
\def\eea{\end{eqnarray}}

\definecolor{mygreen}{rgb}{0.0,0.55,0.3}

\newcommand{\stkout}[1]{\ifmmode\text{\sout{\ensuremath{#1}}}\else\sout{#1}\fi}

\begin{document}
 \title{Tricritical behavior in dynamical phase transitions}
\author{Tal Agranov}
\affiliation{DAMTP, Centre for Mathematical Sciences, University of Cambridge, Wilberforce Road, Cambridge, CB3 0WA}
\author{Michael E. Cates}
\affiliation{DAMTP, Centre for Mathematical Sciences, University of Cambridge, Wilberforce Road, Cambridge, CB3 0WA}
\author{Robert L. Jack}
\affiliation{DAMTP, Centre for Mathematical Sciences, University of Cambridge, Wilberforce Road, Cambridge, CB3 0WA}
\affiliation{Yusuf Hamied Department of Chemistry, University of Cambridge, Lensfield Road, Cambridge CB2 1EW, United Kingdom}

\begin{abstract}
	We identify a new scenario for dynamical phase transitions associated with time-integrated observables occurring in diffusive systems described by the macroscopic fluctuation theory. It is characterized by the pairwise meeting of first- and second-order bias-induced phase transition curves at two tricritical points. We formulate a simple, general criterion for its appearance and derive an exact Landau theory for the tricritical behavior. The scenario is demonstrated in three examples: the simple symmetric exclusion process biased by an activity-related structural observable; the Katz-Lebowitz-Spohn lattice gas model biased by its current; and in an active lattice gas biased by its entropy production. \end{abstract} 

\maketitle

\emph{Introduction} -- 
In non-equilibrium statistical mechanics, theoretical results for simple lattice models have guided understanding of dynamical processes and fluctuations \cite{jona-lasinio_large_1993,kipnis_scaling_2010,lebowitz_gallavotticohen-type_1999,bertini_fluctuations_2001,garrahan_dynamical_2007,derrida_non-equilibrium_2007,lefevre_dynamics_2007,bodineau_long_2008-1,jona-lasinio_fluctuations_2010,krapivsky_large_2014}.  For interacting particle systems, macroscopic fluctuation theory (MFT) \cite{bertini_fluctuations_2001,bertini_macroscopic_2002,bertini_large_2003,bertini_minimum_2004,bertini_stochastic_2007,bertini_macroscopic_2015-1} enables analysis of hydrodynamic scales, exposing behavior independent of microscopic details. Alongside models' typical behavior, MFT predicts rare fluctuations.  For example, it identifies the fluctuation mechanism for time-integrated quantities whereby atypical values of the current \cite{bodineau_current_2004-1,bodineau_distribution_2005,bertini_current_2005,bertini_non_2006,bodineau_cumulants_2007,derrida_current_2009,shpielberg_chatelier_2016,zarfaty_statistics_2016,baek_dynamical_2017-1} or dynamical activity \cite{appert-rolland_universal_2008,lecomte_inactive_2012,jack_hyperuniformity_2015,vanicat_mapping_2021} are sustained over long times. These are examples of large deviations, which have also been analysed numerically \cite{hurtado_spontaneous_2011}, and by other theoretical methods \cite{derrida_exact_1998}. A rich behavior emerges \cite{touchette_large_2009,chetrite_nonequilibrium_2013,jack_ergodicity_2020}, including dynamical phase transitions (DPTs), often involving spontaneous symmetry breaking by these (macroscopically atypical) system trajectories \cite{bertini_current_2005,bertini_non_2006,garrahan_dynamical_2007,bodineau_cumulants_2007,hurtado_spontaneous_2011,lecomte_inactive_2012,jack_hyperuniformity_2015,shpielberg_chatelier_2016,zarfaty_statistics_2016,baek_dynamical_2017-1,dolezal_large_2019,appert-rolland_universal_2008,lecomte_inactive_2012,jack_hyperuniformity_2015,vanicat_mapping_2021}.

DPTs are conceptually intriguing, and also provide practical insight, in part because large deviation analyses relate directly to optimal control theory \cite{bertsekas_dynamic_2017,dupuis_weak_1997,jack_large_2010,jack_effective_2015,chetrite_variational_2015,bertini_macroscopic_2015-1,jack_ergodicity_2020}. Here  rare events are characterized via extra control forces, added to the system dynamics, to make them become typical.
This approach has applications in numerical experiments and for material design \cite{garrahan_dynamical_2007,jack_large_2010,pinchaipat_experimental_2017,abou_activity_2018,tociu_how_2019,nemoto_optimizing_2019,jack_ergodicity_2020,fodor_dissipation_2020}.  In this setting, DPTs signify qualitative changes in the types of control force required.

Several well-studied DPTs occur in the simple symmetric exclusion process (SSEP), with periodic boundary conditions.  Its 
steady states are homogeneous (H), but large deviations towards low activity occur through spatially inhomogeneous (IH) states, while those with large activity exhibit hyperuniformity \cite{jack_hyperuniformity_2015,lecomte_inactive_2012}.  The transition from H to IH spontaneously breaks translational symmetry and is continuous.  In contrast, discontinuous DPTs also arise, in exclusion processes~\cite{baek_dynamical_2017-1} and other models~\cite{garrahan_dynamical_2007}.  

In this work, we explore a new type of dynamical phase behavior for fluctuations of time-integrated quantities, which manifests as a pair of tricritical points. These live on H-IH phase boundaries and  signal a change in character of the H-IH transition, from continuous to discontinuous.  We analyse this scenario using MFT, showing it has a universal status -- occurring generically when simple criteria are met.  We exemplify this with three large-deviation calculations: fluctuations of a structural observable akin to the activity in SSEP; fluctuations of the current in a Katz-Lebowitz-Spohn (KLS) type lattice gas; and fluctuations of the entropy production in an active lattice gas model. 

Note that current fluctuations in 1D have been extensively studied \cite{bodineau_current_2004-1,bertini_current_2005,bertini_non_2006,bodineau_cumulants_2007,derrida_current_2009,shpielberg_chatelier_2016,zarfaty_statistics_2016,baek_dynamical_2017-1}, including recent exact solutions via MFT \cite{bettelheim_inverse_2022,mallick_exact_2022,grabsch_exact_2022}, for cases where the mobility depends quadratically on density.
We show below that tricriticality generically arises when the mobility has an inflection point, absent in those studies, creating a much richer picture for DPTs than previously identified. (The possibility of discontinuous transitions was noted in \cite{bodineau_cumulants_2007}, but tricritical points have not been explored, to our knowledge.)

\emph{Large deviations in SSEP} -- We first address fluctuations of time-integrated structural quantities in the SSEP.  Consider a one-dimensional periodic lattice with $L$ sites and $N$ particles;  each site contains at most one particle, and particles hop to vacant neighbors with rate $D_0$.  To analyse the hydrodynamic scale, let the position of site $i$ be $x=i/L$, and write $\rho(x,t)$ for the hydrodynamic density, with time $t$ measured on the hydrodynamic scale.  (The microscopic time is then $\hat{t}= L^2t$.)  Also write $J(x,t)$ for the hydrodynamic current, and denote by ${\cal X}=\{\rho(x,t),J(x,t)\}_{x\in[0,1),t\in[0,T]}$ a dynamical trajectory of duration $T$.   Such trajectories respect the continuity equation $\partial_t \rho = -\nabla \cdot J$, so the total density of the system, $\rho_0=N/L$, is conserved.  

Within MFT, the probability of a trajectory is $P({\cal X}) \simeq {\rm e}^{-L S({\cal X})} $ \cite{kipnis_scaling_2010,bertini_fluctuations_2001,bertini_macroscopic_2002,bertini_large_2003,bertini_minimum_2004,bertini_stochastic_2007,bertini_macroscopic_2015-1}, with action 
\beq
S_T({\cal X}) =  \int_0^T dt \! \int_0^1 dx\,  \frac{| J + D(\rho) \nabla \rho |^2}{2 \sigma(\rho)} 
\label{equ:action-ssep}
\eeq
where $D(\rho)=D_0$ and $\sigma(\rho)=2D_0\rho(1-\rho)$.  Below we retain $D$ and $\sigma$ as general functions, specialising to SSEP where appropriate.
We consider large deviations of time-integrated structural quantities of the form
\beq
K_T({\cal X}) = L \int_0^T dt \! \int_0^1 dx\, \kappa(\rho) \; , \label{K}
\eeq
with two exemplar choices for $\kappa(\rho)$:
\begin{align}\label{k12}
\kappa_1(\rho) = \rho(1-\rho) \quad,\quad
\kappa_2(\rho) = \rho(1-\rho)^2  \; .
\end{align}
For $\kappa=\kappa_1$, $K_T$ measures the dynamical activity~\cite{lecomte_inactive_2012,jack_hyperuniformity_2015}: it counts the number of possible particle hops (i.e., particles with a vacant neighbor). Meanwhile, $\kappa_2$ counts particles with {\em two} vacant neighbors \cite{Note1}.  Despite their physically similar definitions, these quantities have contrasting large deviation behaviors.

To analyse this, we define the scaled cumulant generating function (CGF) $\Psi(\Lambda) = \lim_{L,T\to\infty} \frac{1}{LT} \log \langle {\rm e}^{\Lambda K_T} \rangle$, where angle brackets indicate a steady-state average. 
Analogous to a thermodynamic potential, the CGF is a `dynamical free energy' for an ensemble of trajectories biased by the field $\Lambda$ conjugate to $K_T$ \cite{touchette_large_2009,jack_large_2010,chetrite_nonequilibrium_2013,jack_ergodicity_2020}. For large $L$, the average is dominated by the most likely trajectory and for large $T$ this is homogeneous in time, so that \cite{Note1}
\beq
-\Psi(\Lambda) = \inf_{\rho\colon\int_0^1 dx \rho  = \rho_0}  \int_0^1 dx \left[  M(\rho)| \nabla \rho |^2 - \Lambda \kappa(\rho) \right]
\label{equ:Psi-var}
\eeq
where $M(\rho)=D(\rho)^2/2\sigma(\rho)$.

An alternative characterization of large deviations involves the rate function $\mathcal I$.  The probability density for $K_T$ obeys for large $L,T$ 
\beq
\log {\rm Prob}[K_T/(LT) \approx k] \simeq  -LT\, \mathcal{I}(k)\,.
\label{equ:rate-hydro}
\eeq  
$\mathcal{I}$ corresponds to a thermodynamic potential dual to $\Psi$, governing an ensemble of trajectories where $K_T$ is fixed.
It can be computed in terms of a dominant path which minimises the action at constrained $K_T$:
\beq
\mathcal{I}(k) = \inf_{{\cal X} \colon K_T({\cal X}) = kLT} S_T({\cal X})/T. 
\label{equ:rate-var}
\eeq
As in thermodynamics, enforcing the constraint by Lagrange multiplier shows that ${\cal I}$ and $\Psi$ are related by Legendre transform. 

\begin{figure}
	\includegraphics[width=4.48cm]{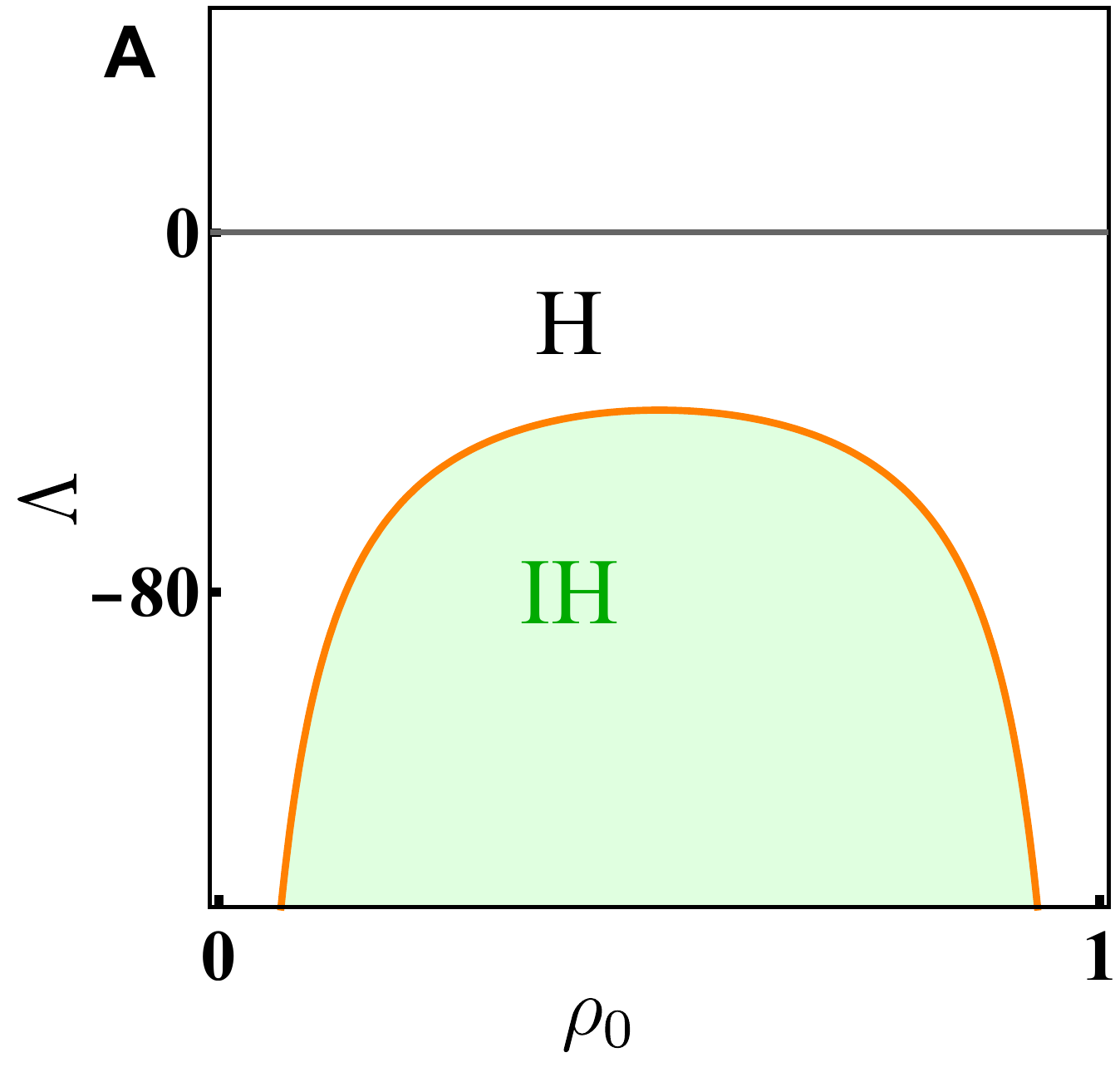}
	\includegraphics[width=4.07cm]{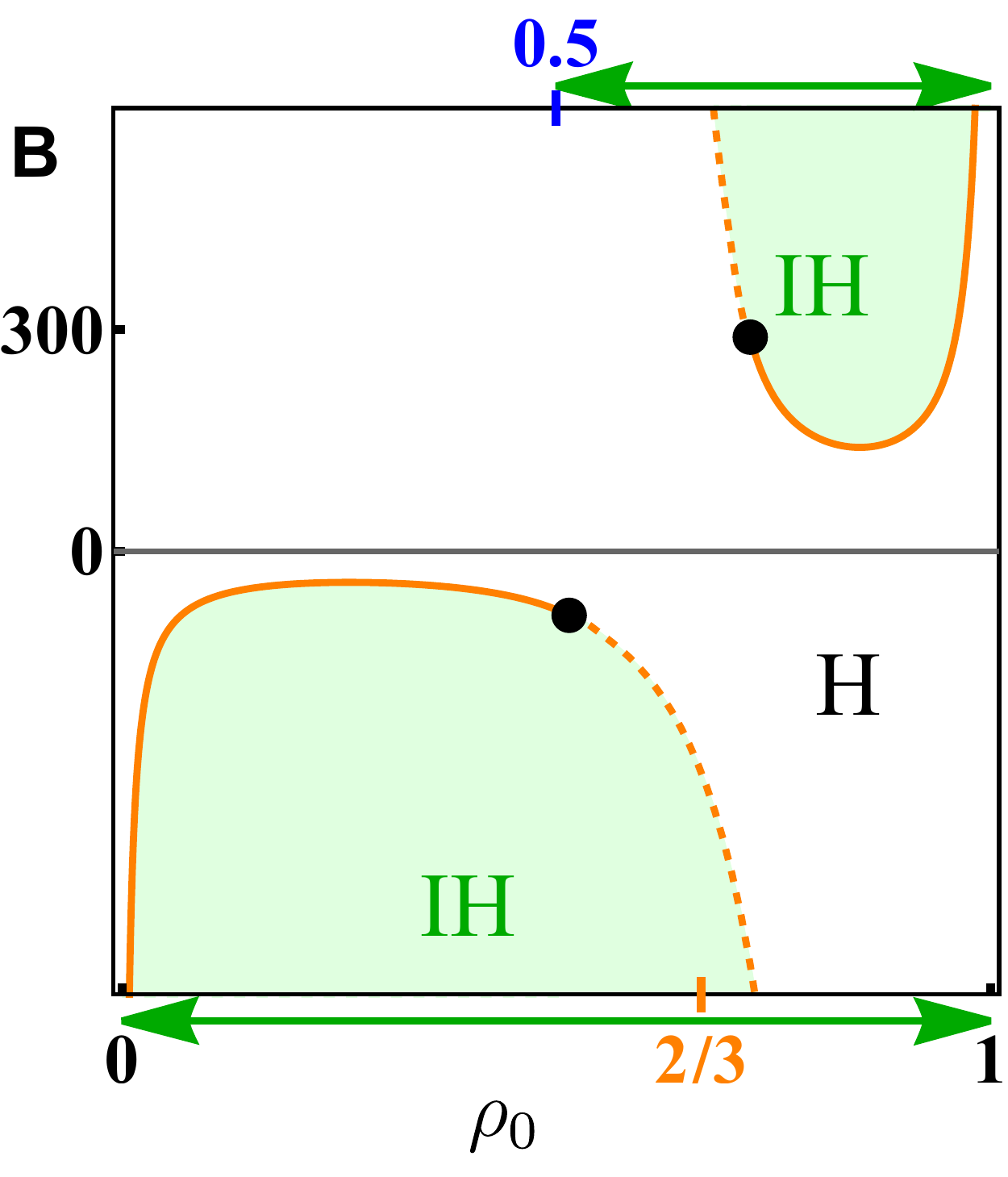}
	\caption{(A,B) SSEP dynamical phase diagrams for (A) $\kappa=\kappa_1$; (B) $\kappa =\kappa_2$, showing H and IH states.
		Arrows demarcate the range of densities $\rho_0$ for which IH states appear as $|\Lambda_c|\to\infty$.
		The thick (orange) lines indicates continuous transitions at $\Lambda=\Lambda_{c,2}$ \eqref{eq:land} and the dashed continuations indicate discontinuous transitions. These meet at tricritical points (black dots).}
	\label{fig:phase-ssep}
\end{figure}

\emph{Dynamical phase transitions in SSEP} -- Fig.~\ref{fig:phase-ssep} shows dynamical phase diagrams for large deviations of $K_T$ in ensembles biased via $\kappa_1$ and $\kappa_2$.  Both cases support H-IH phase transitions, but biasing by $\kappa_2$ introduces tricritical points, absent for $\kappa_1$. To explain this, we first establish a simple condition for discontinuous transitions, related to previous arguments at microscopic level~\cite{garrahan_dynamical_2007,garrahan_first-order_2009,Note1}. This sufficient condition only involves $\kappa(\rho)$, although discontinuous transitions could also arise for sufficiently elaborate choices of $M(\rho)$ \cite{Note1}. 

IH states occur when the minimizer of (\ref{equ:Psi-var}) has $\rho(x)\neq\rho_0$. The latter is optimal for  $\Lambda=0$, whereas for $\Lambda\to-\infty$ the gradient term is negligible and we minimise $\int\kappa(\rho)\,dx$.  The outcome depends on the convexity of $\kappa$: IH profiles are optimal whenever $\kappa(\rho)$ differs from its lower convex envelope, which is the lower boundary of the convex hull. (This condition is analogous to the double tangent construction for thermodynamic phase separation.) The resulting minimiser has two spatial regions, separated by an interface of width $O(|\Lambda|^{-1/2})$. For both $\kappa_{1}$ and $\kappa_2$ these have bulk densities $\rho=0,1$. 

In such cases, the system is IH for $\Lambda\to-\infty$ but H for $\Lambda=0$: clearly there must be an intervening DPT where translational symmetry is broken.  The same argument applies for $\Lambda\to+\infty$, on replacing $\kappa$ by $-\kappa$.  
The arrows in Fig.~\ref{fig:phase-ssep}(B), show the regions of IH for large $|\Lambda|$.  
Only if $\kappa$ has an inflection point (so that neither of $\pm\kappa$ is convex) do IH states exist for both signs of $\Lambda$.

We next establish conditions governing the order of these DPTs.
At a {\em continuous} transition $\rho(x)$ deviates smoothly from $\rho_0$ as bias is increased.
Using \eqref{equ:Psi-var}  with $\Lambda<0$, this requires
a small perturbation to reduce $\int\kappa(\rho)\,dx$, implying $\kappa''(\rho_0)<0$. Conversely, if $\kappa''(\rho_0)>0$ any transition must be discontinuous. 
Summarising: for any $\rho_0$ at which $\kappa$ differs from its lower convex envelope then an H-IH transition must occur for some $\Lambda<0$. If $\kappa''(\rho_0)>0$, then it
{\em must} be first-order; otherwise it may be continuous or discontinuous. (Analogous results again hold for $\Lambda>0$, on replacing $\kappa \to-\kappa$.) 

Since $\kappa_2''(\rho_0)$ changes sign at $\rho_0=2/3$, any H-IH transitions at $\Lambda>0$ is discontinuous for $\rho_0<2/3$; likewise for $\rho_0>2/3$ when $\Lambda<0$. (In fact, the transitions are discontinuous over broader ranges; see below.)  In contrast, $\kappa_1''(\rho_0)<0$ for all $\rho_0$: the H-IH transition is always continuous in that case.

To analyse these DPTs quantitatively,
we develop a Landau theory~\cite{bodineau_cumulants_2007,lecomte_inactive_2012,baek_dynamical_2017-1,dolezal_large_2019}, valid close to tricriticality.  We expand the density as 
$
\rho(x) = \rho_0 + A \cos 2\pi x + B \cos 4\pi x,
$
where $A$ is a small amplitude and $B=O(A^2)$ \cite{Note1}. Substituting into \eqref{equ:Psi-var} yields
\beq\label{eq:land}
-\Psi(\Lambda) \approx -\Lambda \kappa(\rho_0)+\inf_{A}  \left[ \frac{\Lambda_{c,2} - \Lambda}{4} \kappa''(\rho_0) A^2 +  \beta(\rho_0) A^4  \right]
\eeq
where $\approx$ means terms of $O(A^6)$ are omitted; here
\begin{equation}
\Lambda_{c,2}=\frac{8\pi^2 M(\rho_0)}{\kappa''(\rho_0)},
\quad\quad
\beta = \frac{\pi^2M(\rho_0)}{24}\left[3a(\rho_0)-b^2(\rho_0) \right],
\label{lamc-beta}
\end{equation}
with $a=2M''/M-\kappa''''/\kappa''$ and $b =3M'/M -\kappa'''/\kappa''$.

The behavior of the Landau theory \eqref{eq:land} is familiar: if $\beta>0$ there is a continuous transition at $\Lambda_c=\Lambda_{c,2}$ beyond which $A\propto\sqrt{|\Lambda-\Lambda_{c,2}|}$. This happens for the SSEP with $\kappa=\kappa_1$  \cite{lecomte_inactive_2012}.  From \eqref{lamc-beta}, the sign of $\Lambda_{c,2}$  matches that of $\kappa''$, as argued previously.

In contrast, if $\beta<0$, symmetry breaking can only happen discontinuously, as already noted in \cite{bodineau_cumulants_2007}.   Points with $\Lambda=\Lambda_{c,2}$ and $\beta(\rho_0)=0$ are tricritical \cite{griffiths_thermodynamics_1970,griffiths_phase_1975,chaikin_principles_1995}: here the transition changes character from continuous to discontinuous \cite{Note1}.
Note also that wherever $\kappa''(\rho_0)\to 0$, $b^2(\rho_0)\to\infty$.  Hence from \eqref{lamc-beta}, $\beta$ is negative in a range of $\rho$ around any inflection point in $\kappa$, such as the one for $\kappa_2$ at $\rho_0=2/3$ (while generically, as in our examples, staying positive elsewhere). 
The two tricritical points that limit this range are easily identified
since $\Lambda_{c,2}$ and $\beta$ are explicit functions \cite{Note1}; see Fig.~\ref{fig:phase-ssep}(B). 

\begin{figure}
	\includegraphics[width=4.35cm]{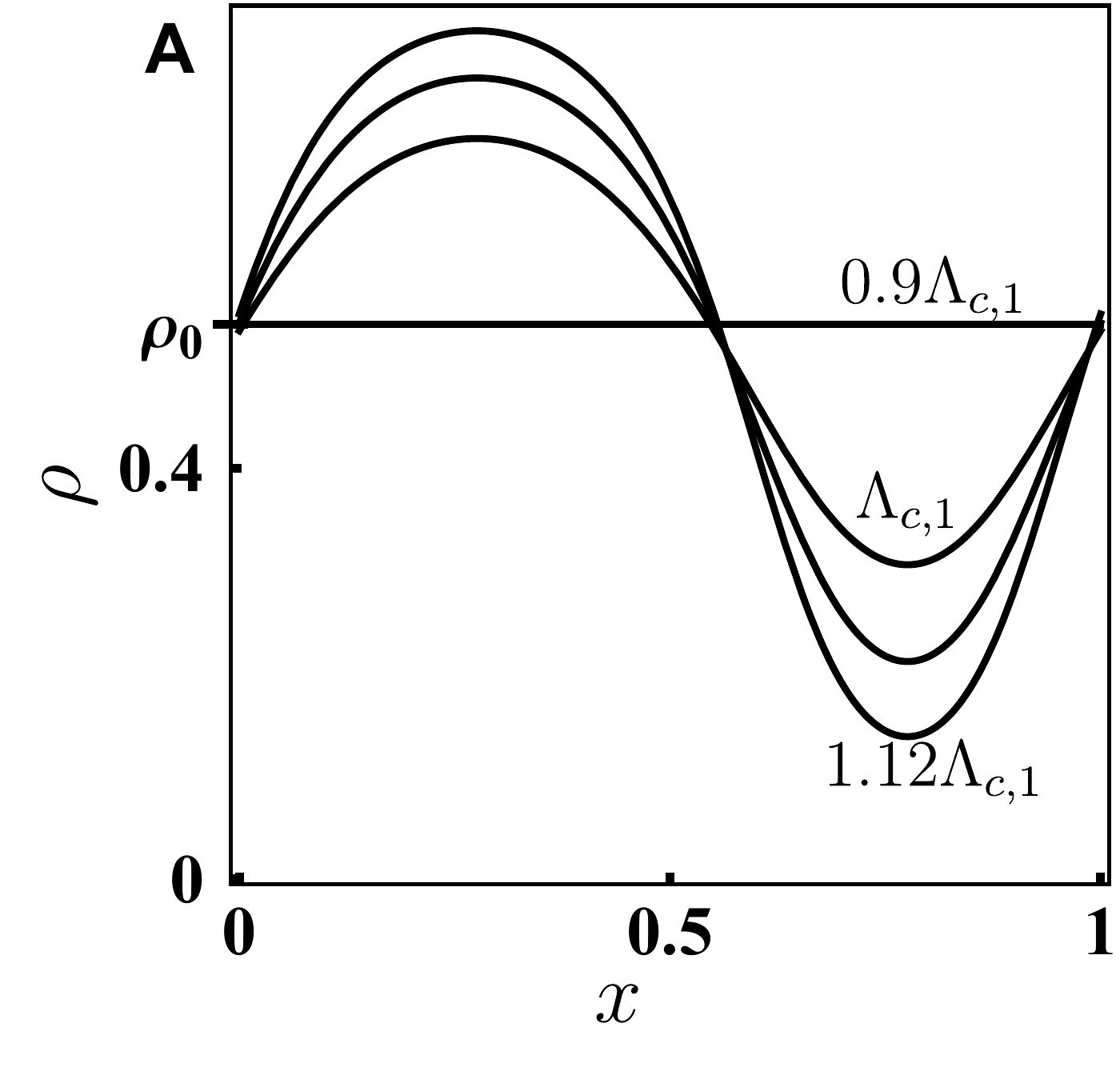}
	\includegraphics[width=4.2cm]{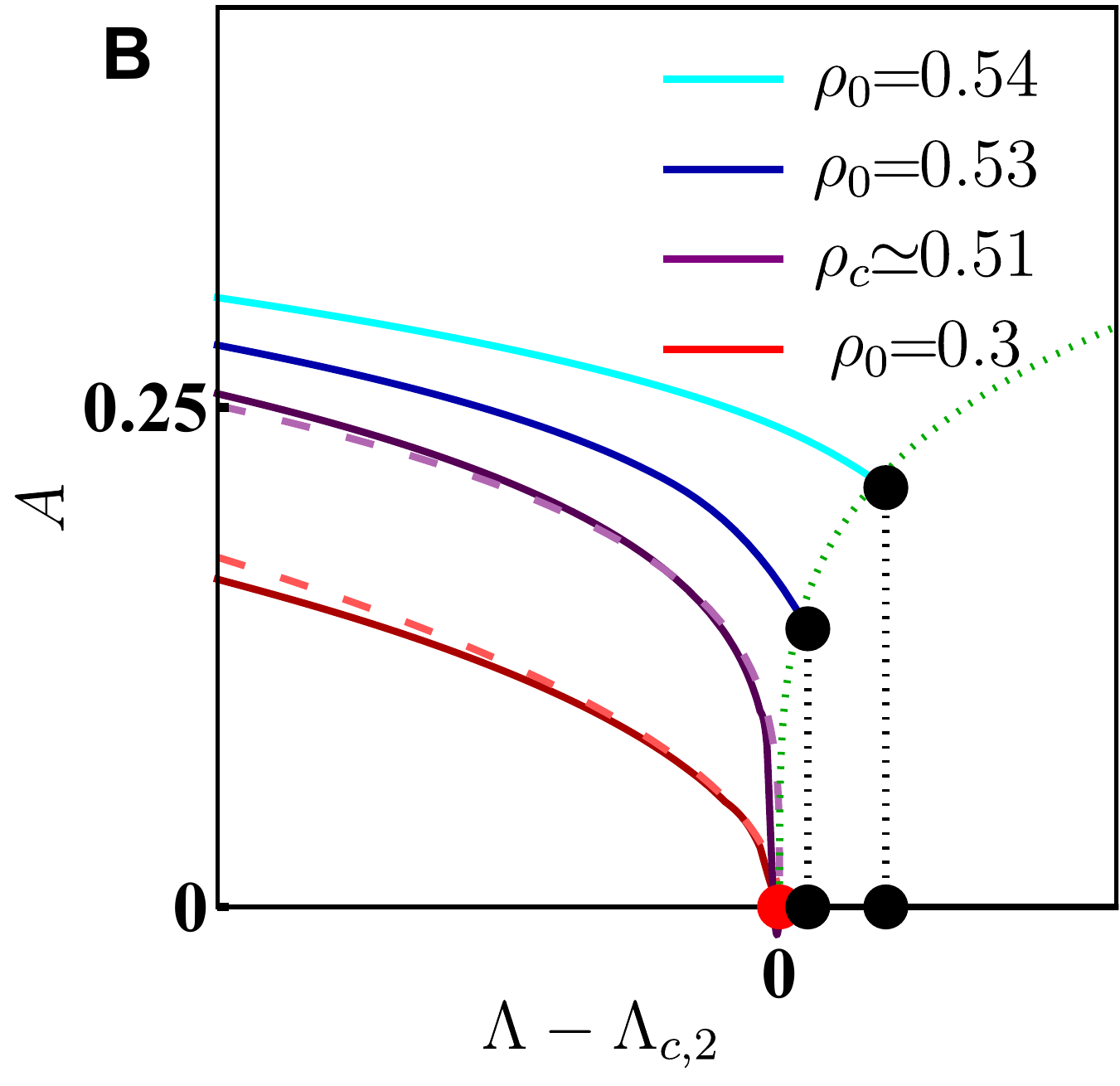}
	\caption{Tricriticality in SSEP for $\kappa=\kappa_2$.  
		(A)~Minimizers of (\ref{equ:Psi-var}), for $\rho_0=0.54$ and $\Lambda=\left(0.9,1,1.11,1.12\right)\times\Lambda_{c,1}$, close to the discontinuous transition at $\Lambda_{c,1}$.  (B)~Amplitude $A$ for various $\rho_0$, near the tricritical point at $(\rho_0,\Lambda)=(\rho_c,\Lambda_{c,2})$.
		For $\rho_0<\rho_c$, a continuous transition occurs at $\Lambda=\Lambda_{c,2}$ where $A\propto|\Lambda_{c,2}-\Lambda|^{1/2}$ (dashed red line) as predicted by \eqref{eq:land}. At $\rho_c\simeq0.515$ the growth follows $A\propto|\Lambda-\Lambda_{c,2}|^{1/4}$ (dashed purple). For $\rho_0>\rho_c$, a discontinuous transition occurs at  $\Lambda=\Lambda_{c,1}$. The discontinuity grows as $\Delta A|_{\Lambda=\Lambda_{c,1}}\propto|\Lambda_{c,2}-\Lambda_{c,1}|^{1/4}$ (dotted green line). Solid lines are numerical solutions of  \eqref{equ:Psi-var}.}
	\label{fig:tri}
\end{figure}

The full tricritical scenario is illustrated in Fig.~\ref{fig:tri} and discussed in \cite{Note1}. If $\beta<0$, and assuming the expansion \eqref{eq:land} is stabilized by a term $\gamma A^6$ with $\gamma>0$, then precisely at the tricritical point, $\beta=0$, one finds $A\propto|\Lambda-\Lambda_{c,2}|^{1/4}$. For $\beta<0$ the transition is discontinuous; it takes place at  $\Lambda=\Lambda_{c,1}$ with $|\Lambda_{c,1}-\Lambda_{c,2}|\propto (\rho_0-\rho_c)^2$. The discontinuity in $A$ grows as $\Delta A|_{\Lambda=\Lambda_{c,1}}\propto|\Lambda_{c,2}-\Lambda_{c,1}|^{1/4}$. These universal, tricritical exponents are exemplified by the theoretical curves in Fig.\ref{fig:tri}(B) which depend on $\gamma$, which we extracted from numerical solutions of \eqref{equ:Psi-var} \cite{Note1}.

\begin{figure}[]
	\begin{tabular}{ll}
		\includegraphics[width=4.1cm]{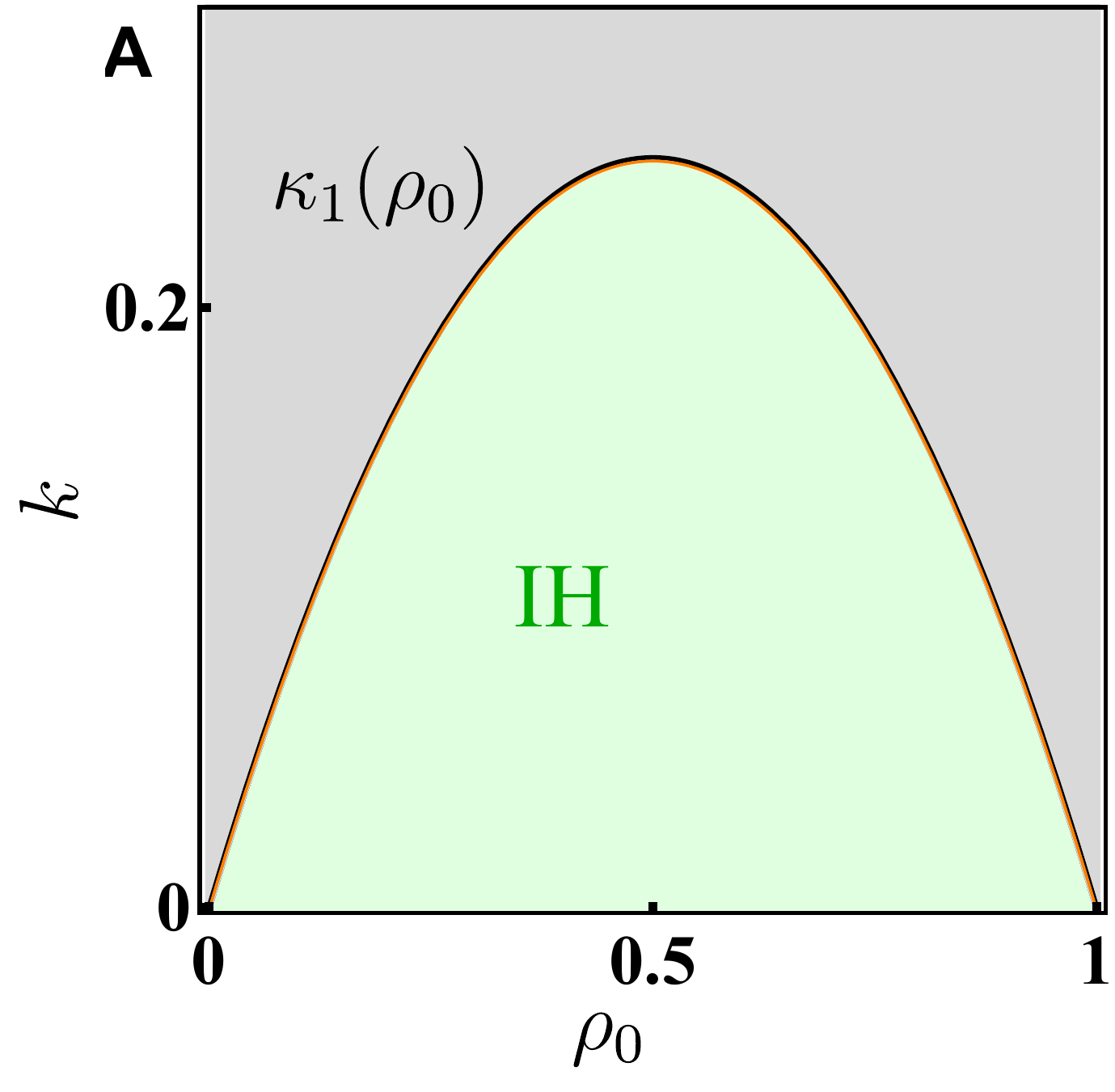}
		\includegraphics[width=3.7cm]{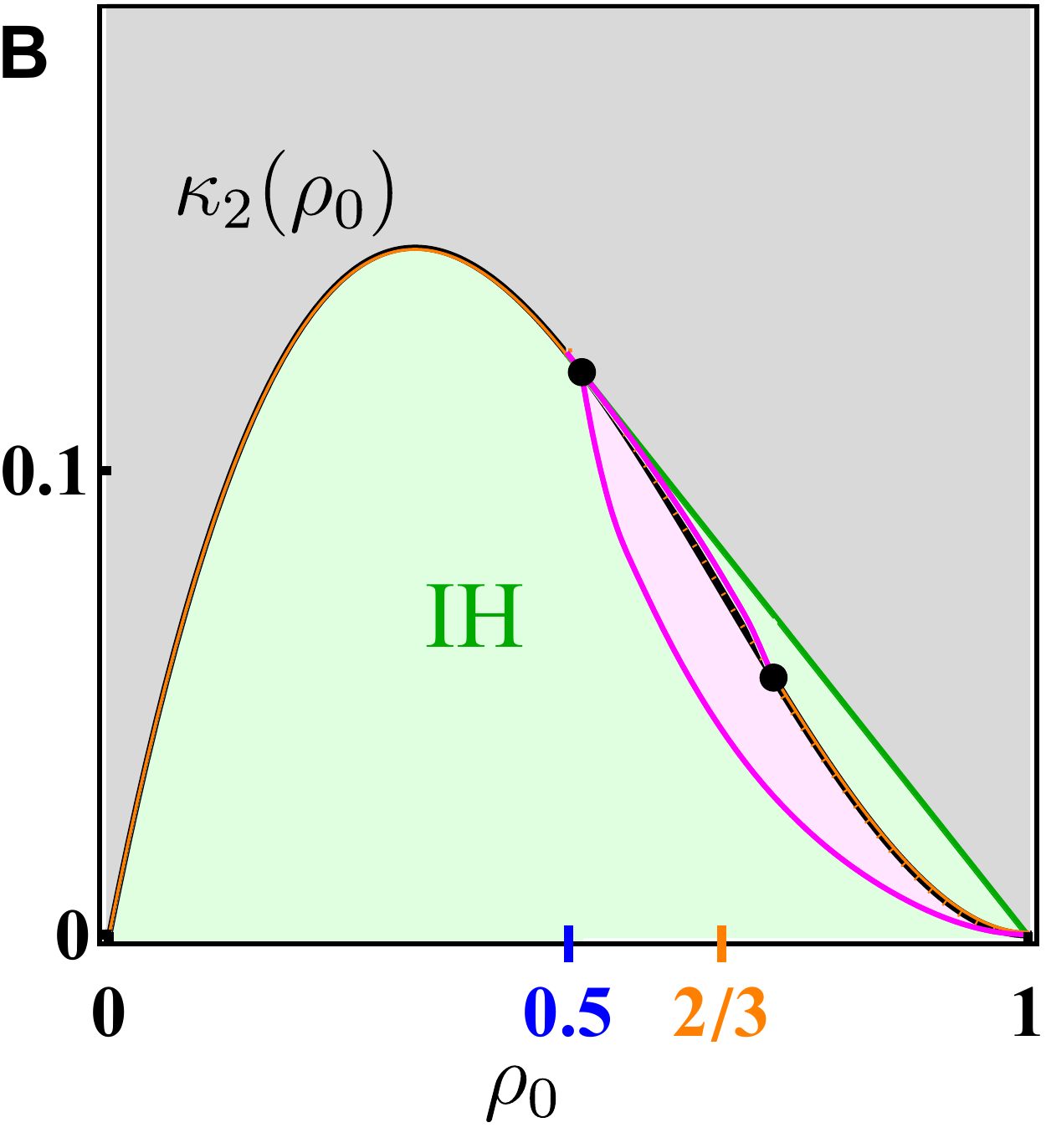}\\
		\includegraphics[width=4.07cm]{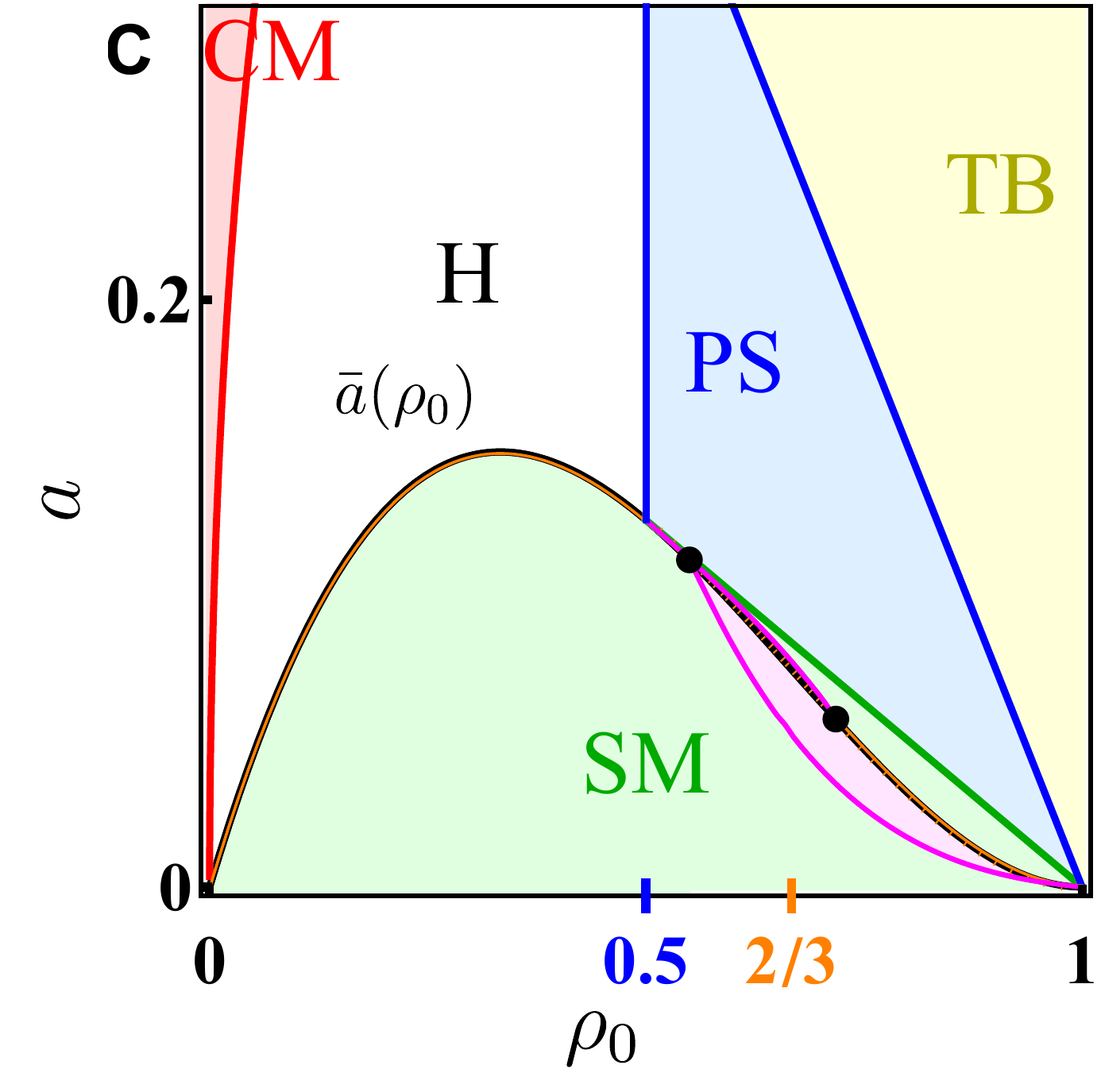}
		\includegraphics[width=4cm]{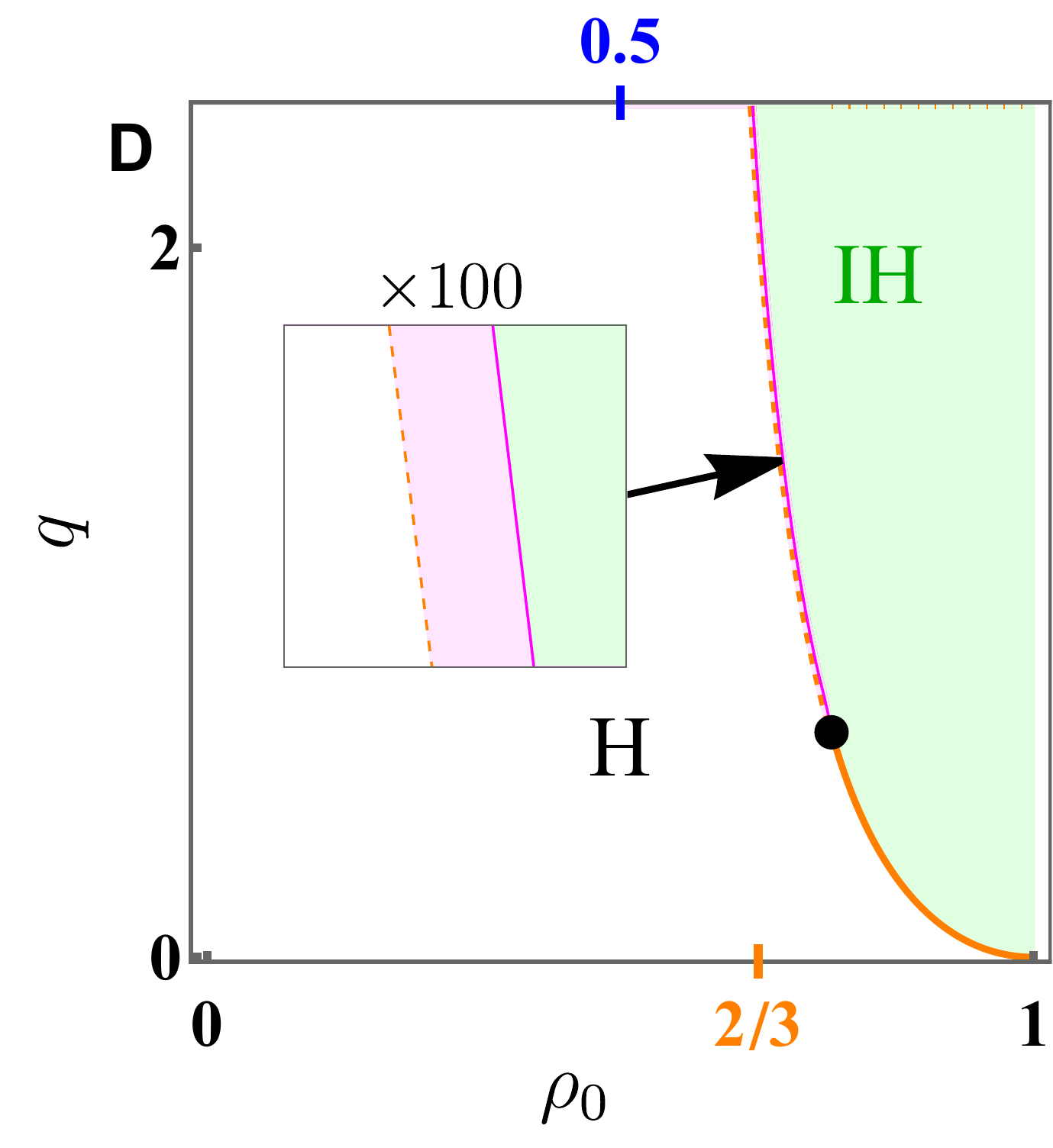}
	\end{tabular}
	\caption{Dynamical phase diagrams for (A)  SSEP conditioned on $\kappa_1$, (B) SSEP conditioned on $\kappa_2$, (C)  active lattice gas conditioned on IEPR, (D)  KLS model conditioned on current.
		Miscibility gaps are denoted by  magenta shading (see inset in (D)). Black dots are tricritical points. Orange tick marks indicate inflection points where $\kappa''=0$ (or $\sigma''=0$ in (D)).  Blue tick marks indicate the boundaries of regions where $-\kappa$ (or $-\sigma$ in (D)) differs from its lower convex envelope. Grey regions in (A,B) are not accessible by hydrodynamic fluctuations.}
	\label{fig:phase-ssep2}
\end{figure}

\emph{Constrained ensemble} -- The variational problem \eqref{equ:Psi-var} is computationally convenient, but additional physical insight is gained via the rate function.
Fig.~\ref{fig:phase-ssep2}(A,B) show dynamical phase diagrams for the constrained ensemble, indicating the fluctuation mechanism, for different values of $K_T$, corresponding to optimal paths in \eqref{equ:rate-var}.   These can be obtained from $\Psi$ by Legendre-Fenchel transform, noting that 
in the presence of first-order DPTs, such optimal paths are inhomogeneous in time \cite{touchette_large_2009,jack_ergodicity_2020,Note1}.  The corresponding regions of `time-like phase separation' (analogous to miscibility gaps in thermodynamics \cite{Note1}) are indicated in Fig.~\ref{fig:phase-ssep2}(B), further highlighting the presence of discontinuous transitions and tricritical points.

When constructing these phase diagrams, it is important that all homogeneous states are identical in MFT, so the entire H phases in Fig.~\ref{fig:phase-ssep}(A,B) collapse onto the lines $k=\kappa(\rho_0)$ in Fig.~\ref{fig:phase-ssep2}(A,B); see also plots in \cite{Note1} showing $\Psi'(\Lambda)=\kappa(\rho_0)$ throughout the H phase.
Physically, this reflects that fluctuations of $K_T$ occur by hydrodynamic mechanisms: the slow relaxation of long-wavelength density modes make their persistent fluctuations much less rare than fluctuations in microscopic structure. However, some values of $k$ are not reached by any hydrodynamic mechanism; in this case the constrained minimisation \eqref{equ:rate-var} has no solution.  
Characterisation of such fluctuations lies beyond MFT (although some aspects of the inaccessible regime can nonetheless be determined~\cite{jack_hyperuniformity_2015,appert-rolland_universal_2008,vanicat_mapping_2021}).

To conclude our study of DPTs in SSEP note that, alongside the emergence of tricritical points, biasing with $\kappa_2$ differs from $\kappa_1$ in that IH states occur for atypical fluctuations at both high and low $\kappa$. At the densities concerned, H states are restricted to a narrow ``tightrope" of unbiased dynamics, $k=\kappa(\rho_0)$. (In contrast, for $\kappa_1$, IH states arise only for low $k$ fluctuations; states at $k>\kappa(\rho_0)$ remain homogeneous \cite{jack_hyperuniformity_2015}.)
We emphasize that this phenomenology should be generic in variational problems like (\ref{equ:Psi-var}), whenever $\kappa$ has a point of inflection. To illustrate this, we now present two further, very different systems where a similar tricritical scenario arises. 

\emph{Current fluctuations} -- We consider large deviations of the integrated current
$Q_T=L\int_0^Tdt\int_0^1dxJ(x,t)$ within MFT. For $L,T\to\infty$, the probability that $Q_T \approx qLT$, as a function of $q$, takes a large deviation form, similar to \eqref{equ:rate-hydro}.  Here though, H-IH transitions involve formation of travelling waves with velocity $V$, so that $\rho=\rho(x-Vt)$ and $J=J(x-Vt)$ \cite{bodineau_distribution_2005,bertini_current_2005,bertini_non_2006,bodineau_cumulants_2007,zarfaty_statistics_2016}.
The rate function for current then satisfies \cite{Note1}
\begin{equation}
\mathcal I(q)=\inf_{{\rho(x)},\alpha}\int_0^1dx\left[M(\rho)|\nabla\rho|^2 +q^2\kappa_J(\rho;\rho_0,\alpha)\right],\label{j}
\end{equation}
with $\kappa_J=\left[1+\alpha(\rho-\rho_0)\right]^2/{(2\sigma(\rho))}$,
where $\alpha=V/q$ is a variational parameter.  This problem is symmetric in $q$, so we now restrict to $q\geq 0$.

The minimisation problem \eqref{j} for $\mathcal I(q)$ is similar to the problem \eqref{equ:Psi-var}, which previously gave the CGF.
Repeating the previous analyses of convexity and the Landau theory yields two analogous results, detailed in \cite{Note1}.  First, as $q\to\infty$, a travelling wave state is found whenever $-\sigma(\rho_0)$ differs from its lower convex envelope.
Second, the quartic term $\beta A^4$ in the corresponding Landau theory has $\beta\to-\infty$ whenever the mobility $\sigma$ has an inflection point, giving tricritical points ($\beta = 0$).  

The mobility $\sigma$ in this problem plays the same role as $\kappa$ did in large deviations of $K_T$ for SSEP.
This correspondence is further exemplified by a model of
Katz-Lebowitz-Spohn type~\cite{katz_nonequilibrium_1984,popkov_steady-state_1999,hager_minimal_2001-1,baek_dynamical_2017-1}, for a kinetically constrained lattice gas \cite{goncalves_hydrodynamic_2009}.  This is a $1d$ simple exclusion process where the hop rates depend on 
the occupancies of neighboring sites as
\begin{eqnarray}\nonumber
0100\overset{D_0}{\longleftrightarrow} 0010\,\,,\,\, 1100\overset{D_0/2}{\longleftrightarrow} 1010\,\,,\,\,0101\overset{D_0/2}{\longleftrightarrow} 0011.
\end{eqnarray}	
The transition $1101 \leftrightarrow 1011$ is kinetically forbidden \cite{berthier_kinetically_2011-1}, but
the hydrodynamic behavior still obeys diffusive MFT with $D(\rho)=D_0(1-\rho)$ and $\sigma=2D_0\rho(1-\rho)^2$ \cite{baek_dynamical_2017-1}. 

The resulting phase diagram shows a tricritical point at $q>0$ (Fig.~\ref{fig:phase-ssep2}(D)) whose partner lies at negative $q$ (not shown).  Since $\sigma(\rho)\propto\kappa_2(\rho)$, this phase diagram resembles 
the upper half of Fig.~\ref{fig:phase-ssep}(B).  Its form  is robust to variations in hop rates \cite{Note1}.

\emph{Active lattice gas} -- Our final example considers a $1d$ active lattice gas (ALG) model \cite{kourbane-houssene_exact_2018}, first introduced to study motility-induced phase separation \cite{cates_motility-induced_2015}.
It comprises two species of diffusing particles, whose hops are biased in opposite directions, with an additional `tumbling' process where particles change species. Its hydrodynamic behavior can be analysed within MFT \cite{agranov_exact_2021,agranov_macroscopic_2022}; the resulting action $S_{\rm act}$ is analogous to (\ref{equ:action-ssep}).  

We discuss here the emergence of tricritical DPTs in large deviations of the informatic entropy production rate (IEPR), which were previously analysed in~\cite{agranov_entropy_2022}. Write ${\cal X}$ for a hydrodynamic trajectory, and let ${\cal X}^R$ be the corresponding time-reversed trajectory.  Then the IEPR, $\mathbb{S}_T \equiv [S_{\rm act}({\cal X}) - S_{\rm act}({\cal X}^R)]/T$ \cite{fodor_irreversibility_2022,obyrne_time_2022}, quantifies time-reversal symmetry breaking at hydrodynamic scales.  Its average is 
$
\langle \mathbb{S}_T \rangle = L s_0 \bar{a}(\rho_0)
$ 
where 
$\bar{a}(\rho) = \rho(1-\rho)^2
$
and $s_0$ is a constant~\cite{agranov_entropy_2022}. 
The IEPR obeys a large deviation principle resembling (\ref{equ:rate-hydro}),
\beq
\log {\rm Prob}[\mathbb{S}_T/(Ls_0) \approx a] \simeq  -LT\,  \mathcal I(a)
\label{equ:rate-active}
\eeq   
where the rate function $\mathcal I(a)$ can be characterised variationally, 
similarly to (\ref{equ:rate-var}). The resulting phase diagram, fully derived in \cite{agranov_entropy_2022}, is shown in Fig.~\ref{fig:phase-ssep2}(C). It is more complex than for the SSEP. As well as the smoothly modulated state (SM) which is analogous to the IH states discussed above, it supports collective motion (CM) and traveling band (TB) states which break the symmetry between species, and a sharply phase-separated (PS) state.   
Nonetheless, the small-$a$ behavior resembles Fig.~\ref{fig:phase-ssep2}(B). 

In the ALG, the dominant fluctuations involve local particle motions that remain typical for the given local density which becomes non-typical. This results in \cite{agranov_entropy_2022}
\beq
\mathcal  I(a) \propto \inf_{\rho(x) \colon a = \int_0^1dx \bar{a}(\rho)} \int dx {\cal M} (\rho)     |\nabla\rho|^2
\label{equ:rho-var-active}
\eeq 
where ${\cal M}$ encodes all the cost arising from inhomogeneities of the density \cite{agranov_entropy_2022}.
Observing that $\bar{a}(\rho_0)=\kappa_2(\rho_0)$, this
variational problem is again similar to (\ref{equ:rate-var}) with $\kappa=\kappa_2$.
As a result, the behavior in the SM state in Fig.~\ref{fig:phase-ssep2}(C) is analogous to the inhomogeneous state in Fig.~\ref{fig:phase-ssep2}(B), including the tricritical points and the time-like phase separation.  
A central result of this Letter is that the tricritical phenomena unexpectedly encountered in~\cite{agranov_entropy_2022} are {\em not} specific to the ALG, instead exemplifying a quite general scenario as explored above.

\emph{Outlook} -- 
We demonstrated a new class of tricritical behavior that occurs in fluctuations of time-integrated observables when the dynamical action has the general structure (\ref{equ:Psi-var}). We gave three examples from the hydrodynamic analysis of large deviations. In all cases, pairs of tricritical points occur on homogeneous-inhomogeneous phase boundaries, separating continuous from discontinuous transitions. Our results significantly enrich the theory of dynamical phase transitions and add to the classes of systems showing tricriticality in  non-equilibrium \cite{antoniazzi_nonequilibrium_2007,keskin_dynamic_2007,marcuzzi_absorbing_2016,jo_tricritical_2020}, for example in fluctuations of instantaneous rather than time-integrated quantities \cite{aminov_singularities_2014}.

The discontinuous transitions in Fig.~\ref{fig:phase-ssep2} show that even if $k$ is close to its mean value, the large-deviation mechanism may differ strongly from the typical (homogeneous) state: for suitable $\rho_0$, time-like phase separation can appear once $k$ deviates from $\kappa_2(\rho_0)$, in either direction.
Alongside aforementioned relevance to optimal control and design \cite{garrahan_dynamical_2007,jack_large_2010,pinchaipat_experimental_2017,abou_activity_2018,tociu_how_2019,nemoto_optimizing_2019,jack_ergodicity_2020,fodor_dissipation_2020}, such transitions should be directly realizable in several experimental settings \cite{baek_dynamical_2017-1}. These include wave transmission in disordered media \cite{pnini_fluctuations_1989,sarma_probing_2014} and mesoscopic electronic transport \cite{sukhorukov_noise_1999,pilgram_stochastic_2003} where, intriguingly, the relevant mobility can show inflection points \cite{tan_measurement_2007}, as required for tricriticality to emerge.

\begin{acknowledgments}We thank Yariv Kafri for useful comments. TA is funded by the  Blavatnik Postdoctoral Fellowship Programme. Work funded in part by the European Research Council under the Horizon 2020 Programme, ERC grant agreement number 740269. MEC was funded by the Royal Society.
\end{acknowledgments}

\vfill
\begin{widetext}

	\newpage
	\section*{Supplemental Material to the paper ``Tricritical behavior in dynamical phase transitions" by T. Agranov,  M. E. Cates, and R. L. Jack}
	
	This supplemental material serves two main purposes. First, we review some previous results that are discussed in the main text.
	They are presented here in a way which is consistent with the notation used in our work, in order to make this study as self-contained as possible. These are Secs.~\ref{0}, \ref{A}, \ref{B}, \ref{C0}, \ref{C},  \ref{E} and \ref{F}.  
	
	In addition, we provide detailed derivations of some of the new results of the main text.  In Sec. \ref{D}, we show how the arguments (in main text) for the fluctuations of $K_T$  can be extended to analyse fluctuations of the current $J_T$.  In Sec.~\ref{G}, we explain the construction of the phase diagram in Fig.~3.  In Sec.~\ref{H} we briefly discuss how our results for existence of discontinuous transitions are related to previous works on glassy systems. This section also mentions the microscopic origin of $\kappa_2$ in Eq.~(3).
	
	\subsection*{Table of contents}
	
	\begin{enumerate}[(I.)]
		{
			\item Time homogeneous optimal path for the biased ensemble.
			\item Landau theory derivation, Eq.~(7).
			\item  Review of tricritical exponents in Fig.~2.
			\item Tricriticality for sufficiently elaborate choice of $M$.
			\item Current fluctuations
			\begin{enumerate}[{V.A}]
				\item Deriving the minimization problem, Eq.~(9).
				\item Predicting phase transitions and tricriticality by adapting the arguments of the main text. 
				\item  Landau theory.
				\item Current fluctuations in the KLS lattice gas.
			\end{enumerate}
			\item Dynamical phase diagram and `time-like phase separation' in Fig.~3.
			\item Relating the variational argument for H-IH transitions to the previous works~\cite{garrahan_dynamical_2007,garrahan_first-order_2009}. 
		}
	\end{enumerate}

	\section{Time homogeneous optimal path for the biased ensemble}\label{0}
	In this section we provide the proof that the optimal profile of the biased ensemble is time independent (the so called additivity principle \cite{bodineau_current_2004-1}).
	
	The optimal path of the biased ensemble minimizes the action of the biased ensemble
	\begin{equation}
	S_T^{\Lambda}({\cal X}) =  \inf_{\rho, J\colon\dot{\rho}=-\nabla\cdot J} \int_0^T dt \! \int_0^1 dx\,  \left[\frac{| J + D(\rho) \nabla \rho |^2}{2 \sigma(\rho)}-\Lambda\kappa(\rho)\right]. 
	\label{bi}
	\end{equation}
	with some prescribed initial condition in time. At long times, the only role of the latter is to set the total mass at all times $\int_0^1dx\rho(x,t)=\rho_0$.
	Expanding the square we have
	\begin{equation}
	\int_0^T\!\int_0^1 dx\, \frac{| J + D(\rho) \nabla \rho |^2}{2 \sigma(\rho)}=\int_0^T\!\int_0^1 dx\,  \left[\frac{J^2}{2 \sigma(\rho)}+M(\rho)| \nabla \rho |^2\right]+\int_0^T\!\int_0^1 dxJ\cdot\nabla\frac{\delta F}{\delta \rho}
	\end{equation}
	where $M=D^2(\rho)/2\sigma(\rho)$ and $F[\rho]=\int dxf(\rho)$ is the free energy of the unbiased dynamics with density $f''(\rho)=D(\rho)/\sigma\left(\rho\right)$, see e.g. \cite{derrida_non-equilibrium_2007}. Integrating by parts this second term, and using the continuity constraint, we have
	\begin{equation}\label{ac}
	S_T^{\Lambda}({\cal X}) =  \inf_{\rho, J\colon\dot{\rho}=-\nabla\cdot J} \left\{\int_0^T dt \! \int_0^1 dx\,  \frac{J^2}{2 \sigma(\rho)}+\int_0^Tdt\!\int_0^1dx\left[M(\rho)| \nabla \rho |^2-\Lambda\kappa(\rho)\right]+ \! \int_0^1 dx\left(F\left[\rho_{t=T}\right]-F\left[\rho_{t=0}\right]\right)\right\}. 
	\end{equation}
	The last term is sub-extensive in time, and can be neglected (apart from un-physical profiles $\rho$ for which the free energy diverges).

	Now for the second term, compare any time in-homogeneous history $\rho(x,t)$, with the optimal time homogeneous one $\rho(x)$, with the same total mass $\rho_0$. At any time instant we have that
	\begin{equation}
	\int_0^1dx\left[M(\rho(x,t))| \nabla \rho(x,t) |^2-\Lambda\kappa(\rho(x,t))\right]\geq \inf_{\rho\colon\int dx\rho(x)=\rho_0}\int_0^1dx\left[M(\rho(x))| \nabla \rho(x) |^2-\Lambda\kappa(\rho(x))\right].
	\end{equation} 
	Lastly, since the first integral in \eqref{ac} is non negative, and vanishes for $J=0$ which corresponds to the time homogeneous solution, we conclude that at long times, the optimal history has no persistent currents and becomes homogeneous in time $[\rho(x),J=0]$.

	\section{Landau theory derivation, Eq.~(7)}\label{A}

	In this section we present the derivation of the Landau expansion Eq.~(7). Such an expansion appeared in several previous works that studied similar second order transitions within the MFT framework \cite{lecomte_inactive_2012,baek_dynamical_2018,dolezal_large_2019,jack_hyperuniformity_2015}. We repeat it here for the convenience of the reader within a consistent notation.
	
	Consider the variational problem 
	\begin{equation}
	-\Psi(\Lambda) = \inf_{\rho\colon\int_0^1 dx \rho  = \rho_0}  \int_0^1 dx \left[  M(\rho)| \nabla \rho |^2 - \Lambda \kappa(\rho) \right]
	\label{equ:Psi-vars}
	\end{equation}
	For $\Lambda=0$, the homogeneous state is a local minimiser in this problem.  By considering small perturbations about this state, we will show that the homogeneous state is no longer a local minimiser if the bias $\Lambda$ is strong enough.  To this end, write
	$\rho(x) = \rho_0 + \delta\rho(x)$ and expand to second order in $\delta\rho$.   Mass conservation requires $\int_0^1 dx \delta\rho(x)=1$  so the question is whether the homogeneous state is a local minimiser of 
	\begin{equation}
	\int_0^1 dx \left[  M_0 \delta\rho'(x)^2 - \Lambda \kappa''_0 \delta\rho(x)^2 \right]
	\end{equation}
	where primes denote derivative with respect to the argument, and the subscript $_0$ denotes evaluating at $\rho_0$, that is, $M_0=M(\rho)\vert_{\rho=\rho_0}$ and $\kappa''_0=d^2\kappa(\rho)/d\rho^2\vert_{\rho=\rho_0}$. 
	Any instability that occurs takes place via the principal mode
	$
	\delta\rho(x) \propto \cos(2\pi x)
	$
	so it is easily verified that the homogeneous state is stable if $\Lambda\kappa''_0 < 8\pi^2 M_0$, as in Eq.~(8).  
	We write
	\begin{equation}\label{lam2}
	\Lambda_{c,2}=\frac{8\pi^2M_0}{\kappa''_0}
	\end{equation}
	for the value of the bias at the instability.  Note that this bias has the same sign as $\kappa''_0$, which may be either positive or negative.
	If $\kappa_0''>0$ then the homogeneous state is unstable for $\Lambda>\Lambda_{c,2}>0$ while for $\kappa_0''<0$ it is unstable for $\Lambda<\Lambda_{c,2}<0$.  We write
	\begin{equation}\label{eps}
	\epsilon=\frac{\Lambda-\Lambda_{c,2}}{\Lambda_{c,2}} \; .
	\end{equation}
	so that the stability criterion is $\epsilon<0$.

	To obtain the behaviour in the unstable regime, we
	expand the density about the homogeneous state as~\cite{lecomte_inactive_2012,dolezal_large_2019}
	\begin{equation}\label{exp}
	\rho(x)=\rho_0+\sqrt{\epsilon}\rho_1(x)+\epsilon\rho_2(x)+\mathcal O(\epsilon^{3/2}).
	\end{equation}
	Since mass conservation must be obeyed at any order of the expansion we have that
	\begin{equation}\label{mass}
	\int_0^1dx\rho_1(x)=\int_0^1dx\rho_2(x)=0.
	\end{equation}
	Also, the linear stability analysis above already indicates that $\rho_1 \propto \cos(2\pi x)$ so it is natural to write $\rho(x) = \rho_0+ A\cos(2\pi x)+\epsilon\rho_2(x)$ with $A={\cal O}({\epsilon}^{1/2})$. Plugging this solution into Eq.~\eqref{equ:Psi-vars} one can show, using mass conservation, integration by parts and the relation \eqref{lam2}, that 
	\begin{eqnarray}\label{sublead}
	&-&\Psi(\Lambda)= -\Lambda_{c,2}\kappa_0(1+\epsilon) \\\nonumber
	&+&\Lambda_{c,2}^2\epsilon^2\inf_{A_0,\rho_2}\int_0^1dx\left\{M_0\left[\rho_2'(x)^2-4\pi^2\rho_2(x)^2\right]+2\pi^2\rho_2(x) A_0^2 \cos(4\pi x)\left(3M'_0-M_0\frac{\kappa_0'''}{\kappa_0''}\right)+\frac{\kappa_0''}{4}A_0^2+A_0^4\pi^2M_0\left(\frac{M_0''}{4M_0}-\frac{\kappa''''_0}{8\kappa''_0}\right)\right\}\\\nonumber&+&\mathcal O(\epsilon^{5/2}).
	\end{eqnarray}
	where $A_0 = \epsilon^{-1/2}A={\cal O}(1)$. Minimizing with respect to $\rho_2$ (which should be orthogonal to $\rho_1$), one finds 
	\begin{equation}
	\rho(x)=\rho_0 + A\cos(2\pi x) + B\cos(4\pi x) \quad \hbox{with} \quad B=-A^2\left(\frac{M_0'}{4M_0}-\frac{1}{12}\frac{\kappa_0'''}{\kappa_0''}\right).
	\end{equation}
	so $B=\mathcal{O}(\epsilon)$.
	Plugging this solution back into \eqref{sublead} and re-expressing the solution in terms of $A,\Lambda$ we arrive at the Landau expansion 
	\begin{equation}\label{min}
	-\Psi(\Lambda) = -\Lambda \kappa(\rho_0)+\inf_{A}  \left[ -2\pi^2M_0\epsilon A^2 +  \beta(\rho_0) A^4  +\mathcal O(A^{6}) \right]
	\end{equation}
	with
	\begin{equation}
	\beta(\rho_0) = \frac{\pi^2M_0}{24}\left[6\frac{M''_0}{M_0}-3\frac{\kappa''''_0}{\kappa''_0}-\left(3\frac{M'_0}{M_0} -\frac{\kappa'''_0}{\kappa''_0}\right)^2 \right],
	\label{lamc-betas}
	\end{equation}
	as reported in Eq.~(7) and Eq.~(8).
	
	Fig.~\ref{beta} shows the function $\beta(\rho)$ for the case of the SSEP, and the observable $\kappa_2(\rho)=\rho(1-\rho)^2$. The roots of $\beta$ at $\rho_{c,1}=0.514\dots$ and $\rho_{c,2}=0.723\dots$ are positioned on opposite sides of the inflection point $\kappa_2''=0$ where $\beta\to-\infty$.	
	
	\begin{figure}[t]
		\includegraphics[scale=0.4]{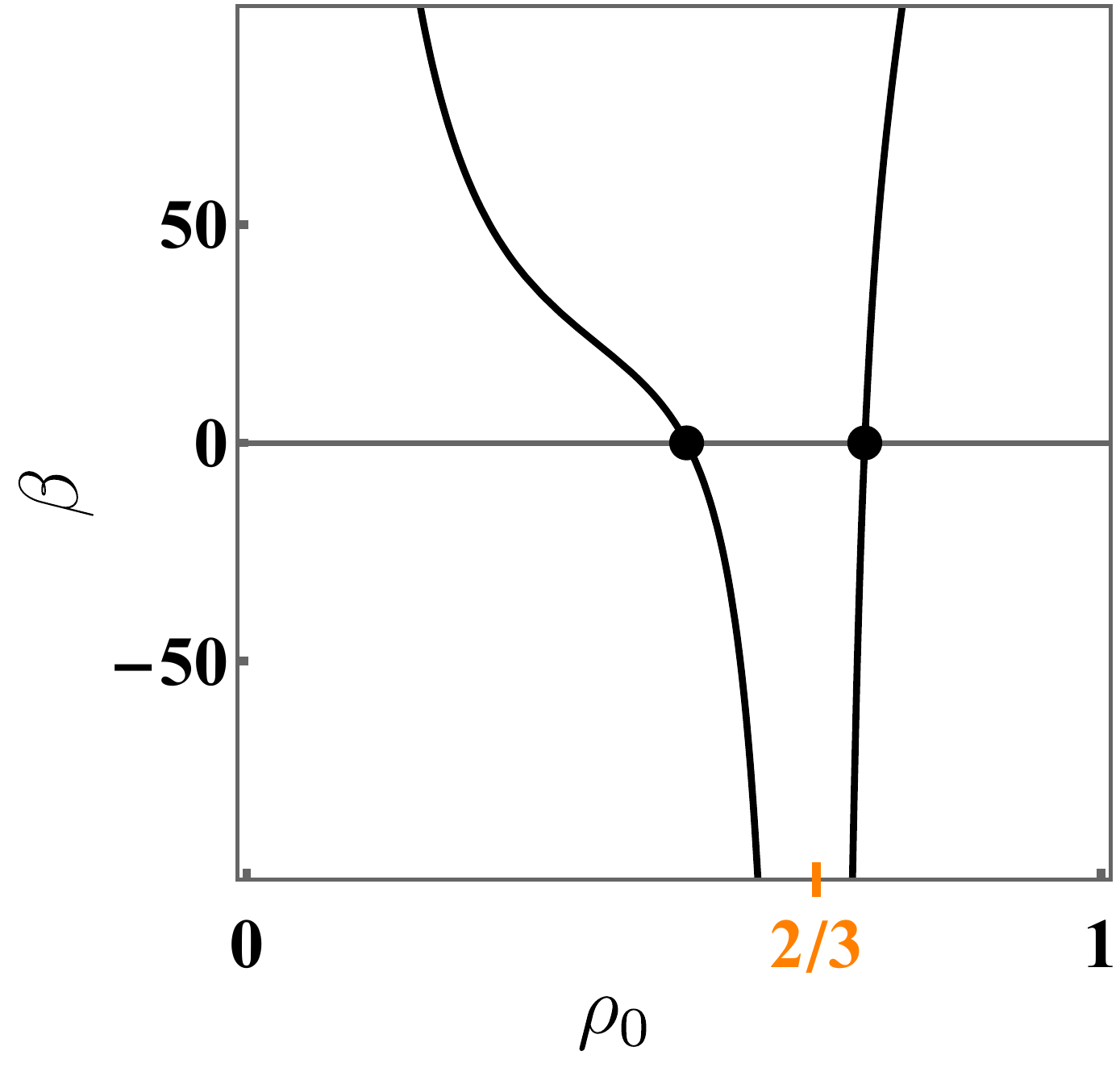}
		\caption{The function $\beta(\rho_0)$ \eqref{lamc-betas} for the case of the SSEP, and the observable $\kappa_2(\rho)=\rho(1-\rho)^2$. The black points are its roots, marking tricriticality. The $2/3$ tick marks the inflection point $\kappa_2''=0$ where $\beta\to-\infty$.}
		\label{beta}	
	\end{figure}

	\section{Review of tricritical exponents in Fig.~2}\label{B}
	In this section we recall the universal exponents describing tricriticality in the Landau expansion \eqref{min}. 
	Their derivation can be found in textbooks on critical phenomena such as \cite{chaikin_principles_1995}.
	
	For $\beta>0$, the minimization \eqref{min} is given by $A=0$ for $\epsilon<0$, while it follows the usual square root growth at positive $\epsilon>0$
	\begin{equation}
	A=\sqrt{\frac{\epsilon\pi^2M_0}{\beta}}
	\end{equation}
	For $\beta<0$, the minimization \eqref{min} must be stabilized by higher order terms. From symmetry these only include  even powers of $A$ and so the minimization reads
	\begin{equation}\label{min2}
	-\Psi(\Lambda) = -\Lambda \kappa(\rho_0)+\inf_{A} \psi_A(A)\quad,\quad \psi_A(A)=  -2\pi^2M_0\epsilon A^2 +  \beta(\rho_0) A^4  +\frac{\gamma}{3} A^6+\mathcal O(A^8).
	\end{equation}
	We will assume $\gamma>0$. We found that this holds true for the case we considered here by direct numerical solutions of the full minimization problem Eq.~\eqref{equ:Psi-vars}, see Fig. \ref{logplot}.
	
	Now consider how the behavior depends on $\epsilon$.
	For $\beta=0$, there is a tricritical point and the transition is still continuous: one has $A=0$ for $\epsilon<0$, while for small positive $\epsilon$ the amplitude follows a modified power law growth
	\begin{equation}\label{a}
	A=\left(\frac{2\pi^2\epsilon M_0}{\gamma}\right)^{1/4}
	\end{equation}
	as reported in the main text and plotted in Fig.~2 (B). By solving the minimization problem \eqref{equ:Psi-vars} numerically, we can then extract the value of $\gamma$ from the plot for $A$, see Fig.~\ref{logplot}. From here we find that for the SSEP biased by $\kappa_2$ this value is $\gamma=290$ (2 sig fig).

	\begin{figure}[t]
		\includegraphics[scale=0.4]{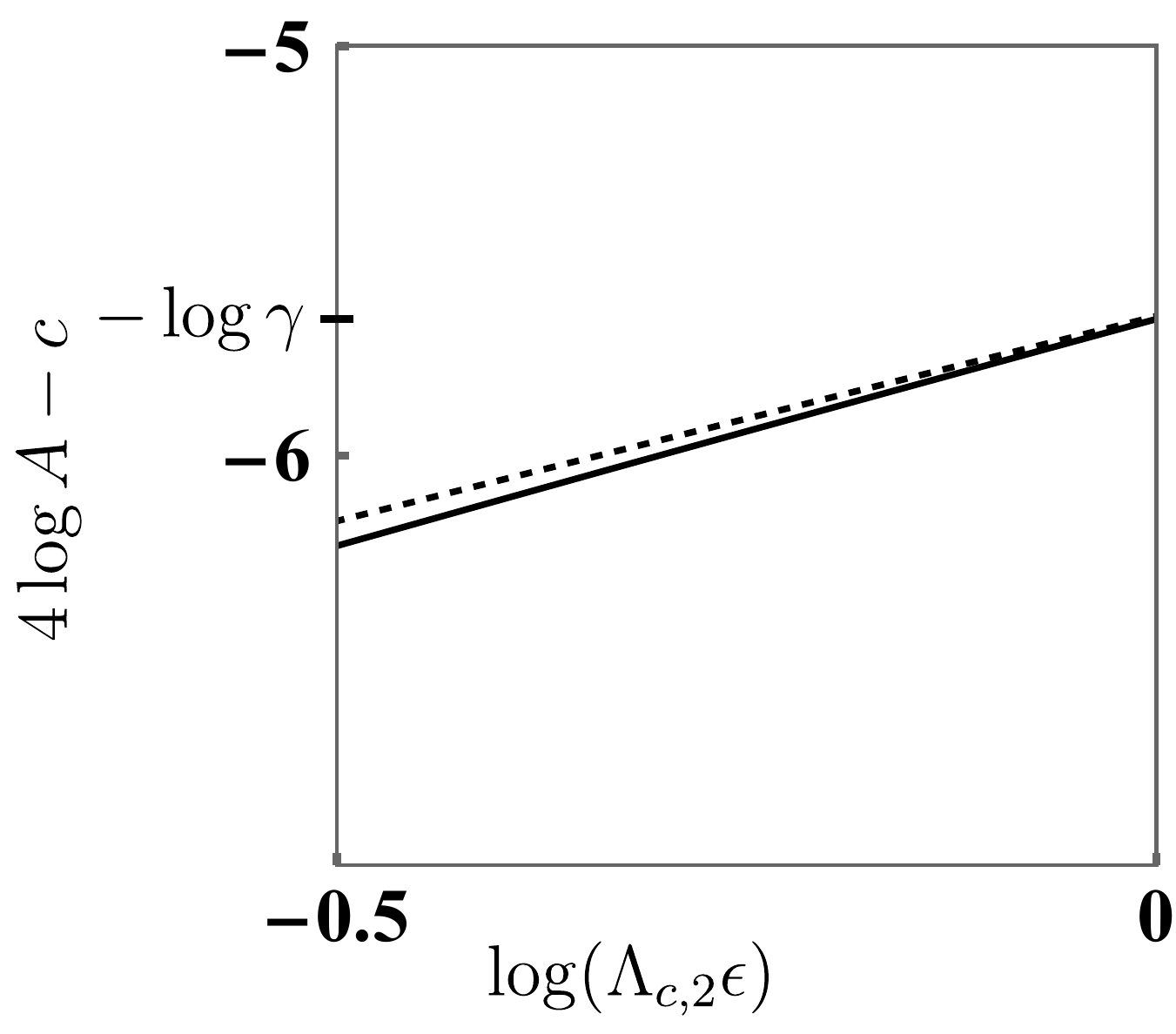}
		\caption{A plot of $4\log A -c$ with $c=\log(-\kappa''_0/4)$ as a function of $\log (\Lambda_{c,2}\epsilon)$ at $\rho_0=\rho_{c,1}$ in thick black line. It was obtained by numerical solution of the full minimization problem \eqref{equ:Psi-vars}, where $A$ is the amplitude of the principal Fourier component $\rho=A\cos(2\pi x)+\dots$. The dashed line indicates the constant slope $1$ as predicted by \eqref{a}. The intersection at $\log (\Lambda_{c,2}\epsilon)=0$ provides the value of $\gamma=290$ (2 sig fig).}
		\label{logplot}	
	\end{figure}
	
	For $\beta<0$, one has $A=0$ for large negative $\epsilon$; this changes discontinuously at
	$\epsilon={-3\beta^2}/(8\pi^2M_0\gamma)<0
	$. 
	Recalling that $\epsilon=(\Lambda-\Lambda_{c,2})/\Lambda_{c,2}$
	the discontinuous transition appears at $\Lambda=\Lambda_{c,1}$ with
	\begin{equation}\label{lam}
	\Lambda_{c,1}=\Lambda_{c,2}\left(1-\frac{3\beta^2}{8\pi^2M_0\gamma}\right)
	\end{equation}
	so that $|\Lambda_{c,1}|<|\Lambda_{c,2}|$: the discontinuous transition occurs at a weaker bias than the second order (spinodal) instability at $\Lambda_{c,2}$.
	As long as $\beta$ has a simple root at $\rho_c$, the difference in the critical biases grows with the distance to the tricritical density as 
	\begin{equation}\label{exp1}
	|\Lambda_{c,1}-\Lambda_{c,2}|\propto (\rho_0-\rho_c)^2,
	\end{equation} 
	as discussed in the main text.
	
	At the critical value $\Lambda=\Lambda_{c,1}$, the jump discontinuity in $A$ reads
	\begin{equation}
	\Delta A=\sqrt{\frac{-3\beta}{2\gamma}}=|\Lambda_{c,2}-\Lambda_{c,1}|^{1/4}\left(\frac{3|\kappa_0''|}{4\gamma}\right)^{1/4},
	\end{equation}
	where in the second equality we plugged the relation \eqref{lam} and used the definition \eqref{lam2}. This curve is plotted in dotted green line in Fig.~2 (B). The thick curves in Fig.~2 (B) were computed from the numerical solution of the minimization \eqref{equ:Psi-vars} where $A$ is the amplitude of the principal mode in the Fourier decomposition $\rho(x)=A\cos(2\pi x)+\dots$.
	
	\section{Tricriticality for sufficiently elaborate choice of $M$}\label{C0}
	In the main text we have derived a sufficient condition for the appearance of tricriticality in the variational problem \eqref{equ:Psi-vars}. This condition only involves the coefficient $\kappa\left(\rho\right)$. Nevertheless, one could have other scenarios for tricriticality involving the coefficient $M(\rho)$. The necessary condition for tricriticality is set by the vanishing of the coefficient $\beta$ \eqref{lamc-betas} in the Landau expansion. For instance, for any model where $M(\rho_0)$ has a degenerate root at $\rho_0$, then $\beta$ must vanishes. We are not aware of a diffusive lattice model where this is the case. However, for the active lattice gas model \cite{agranov_entropy_2022}, this happens to be true. For this model, the equivalent coefficient ${\cal M} (\rho)$ (see Eq.~(11)) has a double root at the MIPS critical point. This is a special property of this model which we do not expect to appear in generic lattice gases.
	
	\section{Current fluctuations}\label{current}
	
	\subsection{Deriving the variational problem Eq.~(9)}\label{C}
	
	We now turn to the analysis of current fluctuations.
	We review here a derivation that appeared in several previous works \cite{bodineau_distribution_2005,bertini_non_2006,bodineau_cumulants_2007,zarfaty_statistics_2016}. We look for solutions to the constrained minimization
	\begin{equation}
	I(q) =  \min_{\rho,J \colon Tq=\int_0^Tdt\int_0^1dxJ} \left[\frac{1}{T} \int_0^T dt \! \int_0^1 dx\,  \frac{| J + D(\rho) \nabla \rho |^2}{2 \sigma(\rho)}\right]
	\label{equ:action-sseps}
	\end{equation}
	in the form of traveling wave solutions
	\begin{equation} \label{travel}
	\rho(x,t)=\rho(x-Vt)\quad,\quad J(x,t)=J(x-Vt).
	\end{equation}
	with velocity $V$.
	In \eqref{equ:action-sseps} we also implicitly assume the conservation equation
	\begin{equation}\label{conser}
	\partial_t\rho=-\nabla\cdot J,
	\end{equation}
	and that the total mass is set to $\rho_0=\int_0^1dx\rho(x,t)$. 
	Then the constraint \eqref{conser}
	together with the constraint on the integrated current enforces the relation
	\begin{equation}\label{eq:j}
	J=q+V(\rho-\rho_0).
	\end{equation}
	Plugging this into the action \eqref{equ:action-sseps}, and using integration by parts for the cross product term, we arrive at the minimization
	\begin{equation}\label{eq:min1}
	I(q)=\inf_{\alpha,\rho(x)\colon\rho_0=\int_0^1dx\rho(x)}\int_0^1dx\left\{M(\rho)| \nabla \rho |^2 +q^2\kappa_J(\rho;\rho_0,\alpha)\right\}\quad,\quad \kappa_J(\rho;\rho_0,\alpha)=\frac{\left[1+\alpha(\rho-\rho_0)\right]^2}{2\sigma(\rho)}
	\end{equation}
	where 
	$
	\alpha=(V/q)
	$ 
	corresponds to a rescaling of the variational parameter $V$, by the integrated current. This is Eq.~(9).
	
	The main difference between the variational problem \eqref{eq:min1} and the one of \eqref{equ:Psi-vars}, is the addition of the variational parameter $\alpha$.
	The minimization with respect to $\alpha$ can be performed explictly to give \cite{bodineau_distribution_2005,bodineau_cumulants_2007}
	\begin{equation}\label{alpha1}
	\alpha_* =-\frac{\int_0^1dx\frac{\rho-\rho_0}{\sigma(\rho)}}{\int_0^1dx\frac{\left(\rho-\rho_0\right)^2}{\sigma(\rho)}}.
	\end{equation} 
	However this makes it a non-trivial functional of the yet undetermined optimal density, which leaves the two problems  \eqref{eq:min1} and \eqref{equ:Psi-vars} distinct. Still, in the next section we will show how one can adapt the same treatment employed for the first problem \eqref{equ:Psi-vars} to \eqref{eq:min1}.
	
	\subsection{Predicting phase transitions and tricriticality by adapting the arguments of the main text}\label{D}
	
	In this section we analyse the minimization \eqref{eq:min1} by adapting the convexity arguments that appeared in the main text. This analysis did not appear in previous works; it is analogous to the similar argument in the main text for the minimisation  \eqref{equ:Psi-vars}.

	\subsubsection{Phase-separated minimizers as $|q|\to\infty$} \label{glob}
	
	Consider the minimization \eqref{eq:min1} in the large current limit $|q|\to\infty$. Then following the same argument presented for \eqref{equ:Psi-vars}, the gradient term in \eqref{eq:min1} becomes negligible and we are left with minimizing the integral $\min_{\alpha,\rho}\int_0^1dx\kappa_J(\rho;\rho_0,\alpha)$. 
	
	As in the main text, the solution to this problem should either be a homogeneous minimizer, or a sharply phase-separated state with coexisting densities $\rho_h,\rho_l$ separated by sharp interfaces of width ${\cal O}(|q|^{-1})$.  The homogeneous case has $\kappa_J=1/(2\sigma(\rho_0))$ so it only remains to compare this with the value for the optimal phase-separated profile.
	For this latter case, one sees from \eqref{alpha1} that
	\begin{equation}\label{alpha4}
	\alpha_*\simeq\frac{\sigma(\rho_h)-\sigma(\rho_l)}{\sigma(\rho_l)(\rho_h-\rho_0)+\sigma(\rho_h)(\rho_0-\rho_l)}.
	\end{equation}
	Plugging this expression into \eqref{eq:min1}, $\kappa_J$ is now an explicit function of the five variables $(\rho,\rho_0,\rho_l,\rho_h)$. 
	
	We seek a phase-separated solution that minimises this $\kappa_J$.  The method is again analogous to the double tangent construction for thermodynamic phase coexistence.  Since the total density is fixed at $\rho_0$, the fraction of the system that is occupied by the low-density phase must be
	\begin{equation}\label{y}
	y=\frac{\rho_h-\rho_0}{\rho_h-\rho_l}.
	\end{equation} 
	(this is the lever rule from thermodynamics).  The search for the phase-separated minimisers then amounts to construction of the lower convex envelope of $\kappa_J$, or to finding a common tangent that touches $\kappa_J$ at $\rho=\rho_l,\rho_h$.  
	Assuming that $\rho_l,\rho_h$ are not extremal densities (such as $\rho=0,1$ in the SSEP), this requires three conditions
	\begin{itemize}
		\item The tangents at $\rho_l,\rho_h$ have the same gradient 
		\begin{equation}
		\left.\frac{\partial\kappa_J}{\partial\rho}\right|_{\rho=\rho_l}=\left.\frac{\partial\kappa_J}{\partial\rho}\right|_{\rho=\rho_h}\label{1}
		\end{equation}
		\item The two tangents are part of a common straight line 
		\begin{equation}
		\label{2}
		\rho_l\left.\frac{\partial\kappa_J}{\partial\rho}\right|_{\rho=\rho_l}-\left.\kappa_J\right|_{\rho=\rho_l}=\rho_h\left.\frac{\partial\kappa_J}{\partial\rho}\right|_{\rho=\rho_h}-\left.\kappa_J\right|_{\rho=\rho_h}
		\end{equation}
		\item The phase-separated profile has a lower value of $\kappa_J$ than the homogeneous solution
		\begin{equation}
		\left.\kappa_J\right|_{\rho=\rho_l}y+\left.\kappa_J\right|_{\rho=\rho_h}(1-y)<\left.\kappa_J\right|_{\rho=\rho_0}
		\label{3}
		\end{equation}
	\end{itemize}
	If one phase, say $\rho_h$, is at an extremal point, the first two conditions are replaced by a single one.\footnote{
		In this case, $\left. (\rho_h-\rho_l)\partial\kappa_J/\partial \rho\right|_{\rho=\rho_l}=\left.\kappa_J\right|_h-\left.\kappa_J\right|_l$. For the KLS model with $\sigma=2D_0\rho(1-\rho)^2$ and $0\leq\rho\leq1$ then indeed the optimal $\rho_h=1$.  The following analysis also carries through in this case.}
	
	Plugging \eqref{alpha4} into \eqref{eq:min1}, an explicit computation shows that the first condition \eqref{1} is met whenever
	\begin{equation}\label{11}
	\sigma'(\rho_h)=\sigma'(\rho_l).
	\end{equation}
	Using this in the second condition \eqref{2} we find
	\begin{equation}\label{22}
	\rho_l\sigma'(\rho_l)-\sigma(\rho_l)=\rho_h\sigma'(\rho_h)-\sigma(\rho_h).
	\end{equation}
	Lastly, using both the relations \eqref{alpha4} and \eqref{y} in \eqref{3}, we find that
	\begin{equation}\label{00}
	\sigma(\rho_0)<\sigma(\rho_l)y+\sigma(\rho_h)(1-y).
	\end{equation}
	The three conditions (\ref{11},\ref{22},\ref{00}) for $\sigma$ mirror exactly (\ref{1},\ref{2},\ref{3}) for $\kappa$, up to a change of sign: they amount to a lower convex envelope (or common tangent) construction on $-\sigma(\rho)$.
	
	To summarize, as $|q|\to\infty$ the optimal solution to \eqref{eq:min1} becomes sharply separated between bulk phases found by a lower convex envelope construction for $-\sigma(\rho)$.  (This mirrors the simpler minimization \eqref{equ:Psi-vars} analysed in the main text.)  As the system is homogeneous at $|q|=0$ there must be an intermediate critical value $q_c$ where a DPT sets in into a state of a traveling density wave. 
	In the following section we establish a condition for this transition to be discontinuous.

	\subsubsection{Connection of discontinuous transitions to local convexity}\label{loc}
	
	As we have seen for the simpler minimization \eqref{equ:Psi-vars}, to establish the existence of a discontinuous transition it is enough to consider small perturbations of $\rho$ around $\rho_0$. We take $\rho_0$ such that $-\sigma(\rho_0)$ differs from its lower convex envelop so that from the previous Sec.~\ref{glob} we know that a DPT sets in at some critical value $|q_c|$. Now if small 
	perturbations about $\rho_0$ only increase the integral of $\kappa_J$ in \eqref{eq:min1}, this DPT must be discontinuous. 
	
	To determine whether this is the case, we first evaluate the parameter $\alpha$ that enters in \eqref{eq:min1}, for small density modulations, $\rho(x) = \rho_0 + \delta\rho(x)$. Following~\cite{bodineau_distribution_2005,bodineau_cumulants_2007}, one obtains from \eqref{alpha1} that
	\begin{equation}\label{alpha2}
	\alpha_*=\frac{\sigma_0'}{\sigma_0}+\mathcal O(\delta\rho ).  
	\end{equation} 
	Using this value in \eqref{eq:min1}, $\kappa_J$ again becomes an explicit function of the density profile. Thus, a small variation of $\rho$ about $\rho_0$ will increase the integral over $\kappa_J$ whenever $\kappa_J(\rho;\alpha=\frac{\sigma_0'}{\sigma_0},\rho_0)$ is a locally convex function of $\rho$ at $\rho_0$.  Observing that
	\begin{equation}\label{dis}
	\left.\frac{\partial^2_\rho \kappa_J(\rho;\alpha=\frac{\sigma_0'}{\sigma_0},\rho_0)}{\partial\rho^2}\right|_{\rho=\rho_0}=-\frac{1}{2}\frac{\sigma''(\rho_0)}{\sigma^2(\rho_0)}
	\end{equation}
	one sees that $\kappa_J$ is locally convex if and only if $\sigma''(\rho_0)<0$.
	
	Combining this with the results of the previous Section, we finally arrive at following conclusion: \\
	Whenever $-\sigma(\rho)$ differs from its lower convex envelope a dynamical phase transition sets in at some critical current $|q_c|$, and this transition is bound to be discontinuous in the range of densities $\rho_0$ where $\sigma''(\rho_0)>0$. In order that $-\sigma(\rho_0)$  differs from its lower convex envelope, while still having  $\sigma''(\rho_0)>0$, it must have an inflection point: $\sigma''(\rho_*)=0$ for some $\rho_*$.

	\subsection{The Landau theory for current fluctuations}\label{E}
	
	In this section we show how the variational formula \eqref{eq:min1} for current fluctuations can be converted to a Landau theory for the amplitude $A$ of a travelling wave of the form \eqref{travel}.  The result of this computation was previously derived in  \cite{bodineau_cumulants_2007}.  Here we present an alternative derivation that highlights the similarities between the variational problems \eqref{eq:min1} and \eqref{equ:Psi-vars}.  (Note that \eqref{equ:Psi-vars} gives the CGF for fluctuations of $K_T$ while \eqref{eq:min1} gives the rate function for fluctuations of $J_T$, so the physical content of these formulae is quite different.  It is their mathematical structures that are analogous.)
	
	The Landau theory applies to small density modulations of the homogeneous state, that is $\rho=\rho_0 + \delta\rho(x)$.  In that case, it was already shown in Sec.~\ref{loc} that 
	\begin{equation}\label{alpha3}
	\alpha_*=\alpha_0+\delta\alpha \;,\quad\hbox{with}\quad\alpha_0=\frac{\sigma_0'}{\sigma_0} \quad\hbox{and}\quad \delta\alpha = {\cal O}(\delta\rho) \; .     
	\end{equation} 
	Inserting this into the definition for $\kappa_J$ in  \eqref{eq:min1} yields
	\begin{equation}
	\label{expen}
	\kappa_J(\rho(x);\rho_0,\alpha_*)= \kappa_0(\rho(x);\rho_0) + \frac{ \delta\alpha \delta\rho(x) [ 1+ \alpha_0\delta\rho(x)] 
		+  \delta\rho(x)^2 \delta\alpha^2}{2\sigma(\rho(x))}  
	\end{equation}
	with
	\begin{equation}
	\kappa_0(\rho;\rho_0) = \frac{1}{2\sigma(\rho)}\left[1+\frac{\sigma_0'}{\sigma_0}(\rho-\rho_0)\right]^2
	\end{equation}
	From \eqref{expen}, it can additionally be shown that 
	\begin{equation}
	\int_0^1 dx [ \kappa_J(\rho(x);\rho_0,\alpha_*) - \kappa_0(\rho(x);\rho_0) ] = {\cal O}(\delta\rho^4)  
	\end{equation}
	where we used $\int_0^1 dx [\rho(x)-\rho_0]=0$ together with the definition of $\alpha_0$ and a suitable Taylor expansion of $1/\sigma(\rho(x))$.
	Hence,  plugging \eqref{expen} into \eqref{eq:min1} and using that $\alpha_*$ solves the minimization over $\alpha$, we obtain
	\begin{equation}\label{eq:min2}
	I(q)=\inf_{\rho\colon\rho_0=\int_0^1dx\rho(x)} \left[ \int_0^1dx\left[M(\rho)\rho'^2 +q^2\kappa_0(\rho;\rho_0)\right]+\mathcal O(\delta\rho^4)
	\right].
	\end{equation}
	By analogy with the stability analysis of \eqref{equ:Psi-vars}, the homogeneous state of this system becomes unstable for $q^2>q_c^2$ with
	\begin{equation}\label{qc}
	q_c^2=-\frac{8\pi^2M_0}{\left.\frac{\partial^2\kappa_0(\rho;\rho_0)}{\partial\rho^2}\right|_{\rho=\rho_0}}=\frac{16\pi^2M_0\sigma_0^2}{\sigma''_0}.
	\end{equation}
	which is analogous to $\Lambda_{c,2}$ in \eqref{lam2}.  
	Note however that while $\Lambda_{c,2}$ might be either positive or negative, $q^2=q_c^2$ can only be achieved if $q_c^2>0$, so this theory only supports critical points for $\sigma''>0$.  The instability of the homogeneous state occurs via the principal mode.
	
	By analogy with Sec.~\ref{A} we now assume that $\rho(x) = \rho_0 + A \cos 2\pi x + B \cos 4\pi x$ with $B={\cal O}(A^2)$.  
	In this case we have from \eqref{alpha1} that $\alpha_* = \alpha_0 + {\cal O}(A^2)$ and hence $\int_0^1 (\kappa_J - \kappa_0 ) dx= {\cal O}(A^6)$.  Then the integrand of \eqref{eq:min2} becomes
	$ [M(\rho)\rho'^2 +q^2\kappa_0(\rho;\rho_0)]+\mathcal O(A^6) $.
	The Landau expansion of $M$ and $\kappa_0$ in powers of $A$ follows Sec.~\ref{A}. The relevant coefficients are then obtained from (\ref{min},\ref{lamc-betas}); they require evaluation of various derivatives of $\kappa_0$. We find 
	\begin{equation}
	\left.\frac{\partial^2\kappa_0}{\partial\rho^2}\right\vert_{\rho=\rho_0}=-\frac{1}{2}\frac{\sigma''(\rho_0)}{\sigma^2(\rho_0)}\quad,\quad\left.\frac{\partial^3\kappa_0}{\partial\rho^3}\right\vert_{\rho=\rho_0}=-\frac{1}{2}\frac{\sigma'''(\rho_0)}{\sigma^2(\rho_0)}\quad,\quad\left.\frac{\partial^4\kappa_0}{\partial\rho^4}\right\vert_{\rho=\rho_0}=-\frac{1}{2}\frac{\sigma''''(\rho_0)}{\sigma^2(\rho_0)}+3\frac{\sigma''^2}{\sigma^3}.
	\end{equation} 
	Then \eqref{eq:min2} becomes
	\begin{equation}\label{min3}
	I(q) = q^2 \kappa_0(\rho_0)+\inf_{A}  \left[ -\frac{(q^2-q_c^2)}{8} \frac{\sigma_0''}{\sigma_0^2} A^2 +  \beta(\rho_0) A^4  + {\cal O}(A^6) \right]
	\end{equation}
	with
	\begin{equation}
	\beta(\rho_0) = \frac{\pi^2M(\rho_0)}{24}\left[\frac{6M''(\rho_0)}{M(\rho_0)}-\frac{3\sigma''''(\rho_0)}{\sigma''(\rho_0)}+18\frac{\sigma''(\rho_0)}{\sigma(\rho_0)}-\left(3\frac{M'(\rho_0)}{M(\rho_0)} -\frac{\sigma'''(\rho_0)}{\sigma''(\rho_0)}\right)^2 \right].
	\label{eq:beta2}
	\end{equation}
	One can show that this expression is in agreement with the analyses of ref.~\cite{bodineau_cumulants_2007}, although the comparison involves some lengthy algebra. 
	In addition, the expression \eqref{eq:beta2} almost coincides with \eqref{lamc-betas} under $\kappa(\rho)\to\sigma(\rho)$, the only difference being the third term in \eqref{eq:beta2}. They key point is that $\beta$ in \eqref{eq:beta2} must become negative in the vicinity of an inflection point $\sigma''=0$; this is directly analogous to the behaviour of \eqref{lamc-betas} when $\kappa''\approx0$.

	\subsection{Current fluctuations in the KLS lattice gas}\label{F}
	A general form of one-dimensional KLS lattice gas is given in terms of hopping rates \cite{katz_nonequilibrium_1984,popkov_steady-state_1999,hager_minimal_2001-1,baek_dynamical_2017-1}
	\begin{eqnarray}\nonumber
	0100&\overset{D_0(1+\delta)/2}{\longleftrightarrow}& 0010\quad,\quad 1101\overset{D_0(1-\delta)/2}{\longleftrightarrow} 1011,\\
	1100&\overset{D_0(1+\epsilon)/2}{\longleftrightarrow}& 1010\quad,\quad 0101\overset{D_0(1-\epsilon)/2}{\longleftrightarrow} 0011,
	\end{eqnarray}
	where $|\delta|,|\epsilon|<1$. The expression for the corresponding gas coefficients $D(\rho)$ and $\sigma(\rho)$ can be found in \cite{baek_dynamical_2017-1}. Importantly for our discussion, $\sigma(\rho)$ has an inflection point for a suitable range of the parameters $\epsilon$ and $\delta$ \cite{baek_dynamical_2017-1}. Correspondingly, tricriticality occurs over throughouht this range.
	
	For concreteness and to make connection with the other examples presented in the paper we set $\epsilon=0,\delta=1$ for which $D(\rho)=D_0(1-\rho)$ and $\sigma(\rho)=2D_0\rho(1-\rho)^2$ \cite{baek_dynamical_2017-1}. The resulting expression for $\beta$ has a pair of roots on both sides of the inflection point $\sigma''(\rho=2/3)=0$. 
	
	Recall that for current fluctuations, tricriticality is only possible in regions of local convexity $\sigma''>0$, see discussion below Eq.\eqref{qc}. Thus, of the two roots, the only relevant one which marks a tricritical point lies in the region of local convexity $\rho_0\in(2/3,1]$. This point is found to be $(\rho_{c}\simeq0.755,q_c\simeq0.634)$ and is denoted in Fig.~3 (D) of the main text (the twin tricritical point at $(\rho_{c},-q_c)$ is not shown).

	\section{Dynamical phase diagram and `time-like phase separation in Fig.~3}\label{G}	
	
	In this section we recall the transformation from the biased ensemble $(\Lambda,\rho_0)$ to the constrained ensemble $(k,\rho_0)$. We then show how to use it to construct the phase diagram Fig.~3, and establish the miscibility gap. A similar analyses can be found in \cite{garrahan_first-order_2009}.
	
	\subsection{Change of ensembles}
	
	The transformation from the biasing parameter $\Lambda$ to the constrained observable $k$ relies on an ensemble equivalence, akin to that of equilibrium thermodynamics, which is well established within large deviation theory \cite{touchette_large_2009,garrahan_first-order_2009,jack_large_2010,chetrite_nonequilibrium_2013}. The CGF $\Psi(\Lambda)$ serves as a thermodynamic potential of an ensemble of trajectories that are biased by their structural observable $K_T$.
	The probability of trajectory ${\cal X}$ within this ensemble is
	\begin{equation}\label{bias}
	P_{\Lambda}({\cal X})=\frac{e^{\Lambda K_T(\mathcal {\cal X})}P(\mathcal {\cal X})}{Z_\Lambda},
	\end{equation}
	where $P(\mathcal {\cal X})$ is the corresponding unbiased probability, under the stochastic dynamics of the model. The normalization is 
	$
	Z_\Lambda=\int d{\cal X}e^{\Lambda K_T(\mathcal {\cal X})}P(\mathcal {\cal X}) = \langle e^{\Lambda K_T(\mathcal {\cal X})} \rangle
	$.
	Since the definition of the CGF is 
	\begin{equation}
	\label{scgf-def}
	\Psi(\Lambda) = \lim_{L,T\to\infty} (LT)^{-1}  \log \langle e^{\Lambda K_T(\mathcal {\cal X})} \rangle
	\end{equation}
	one has for large $L,T$ that $Z_\Lambda\sim e^{LT\Psi(\Lambda)}$.  Moreover, 
	the expectation of $k=K_T/LT$ with respect to \eqref{bias} behaves for large $L,T$ as
	\begin{equation}\label{kl}
	\langle k\rangle_{\Lambda}=\Psi'(\Lambda).
	\end{equation}
	Similarly, define the probability distribution for trajectories in the constrained ensemble (with $K_T=kLT$): 
	\begin{equation}\label{const}
	P_{k}(\mathcal {\cal X})=\frac{P(\mathcal {\cal X}) \delta(K_T - kLT)}{P_k(k)},
	\end{equation}
	where $P(k) = \int d{\cal X} P(\mathcal {\cal X}) \delta(K_T - kLT)$ is a (rescaled) probability density function for $k$.  For large $L,T$,
	\begin{equation}
	P(k)\sim e^{-LT\, \mathcal{I}(k)} ,
	\end{equation} 
	where ${\cal I}$ is the rate function~\cite{touchette_large_2009}.
	
	By analogy with the equivalence of ensembles in thermodynamics, one has that for large $L,T$, trajectories in the biased ensemble \eqref{bias} at a given value of $\Lambda$ are representative of the trajectories in a constrained ensemble \eqref{const}, for some appropriate value of $k$.  This value is $k=\langle k\rangle_{\Lambda}$ as defined in \eqref{kl}.  
	
	In principle, one should therefore solve $k=\Psi'(\Lambda)$, to obtain the value of $\Lambda$ that corresponds to a constrained ensemble with some given $k$.   The function $\Psi'$ is non-decreasing.  If $\Psi'$ is continuous then the relationship between biased and constrained ensembles is straightforward, and $\Psi$ and $\cal I$ are related by Legendre transform.  
	However, if $\Psi'(\Lambda)$  has a discontinuity at some $\Lambda_*$ where it jumps between two values $k_1$ and $k_2$, then there will be no $\Lambda$ that achieves $k_1 < \Psi'(\Lambda) < k_2$.  Hence, representative trajectories for the constrained ensemble with $k\in(k_1,k_2)$ cannot be obtained by mapping to a biased ensemble.  The generic result for such cases is that
	representative trajectories of the constrained ensemble $P_k$ exhibit time-like phase separation: each trajectory has two parts, which separately resemble trajectories of biased ensembles with $\Lambda = \Lambda_*^\pm$, which have $k=k_1,k_2$.  The division of the total duration $T$ into the two parts is given by the usual lever rule of thermodynamics.  We refer to the range of $k$ between $k_1$ and $k_2$ as a miscibility gap (it is also known as a regime of time-like phase coexistence).

	\subsection{Building the phase diagram of the constrained ensemble}

	We describe how the phase diagrams in the $(\rho_0,k)$ plane are constructed in practice.
	Our discussion is general for any observable $K_T$. 
	As an accompanying example, we consider $K_T({\cal X}) = L \int_0^T dt \! \int_0^1 dx\, \kappa_{2.5}(\rho)$ with 
	\begin{equation}\label{k3}
	\kappa_{2.5}(\rho)=\rho(1-\rho)^{2.5}.
	\end{equation}
	The advantage of this $\kappa_{2.5}$, over $\kappa_2$ which was presented in the main text is that the features of the positive $K_T$ fluctuations are better resolved graphically, see Fig.~\ref{scheme} below. Apart from that, the dynamical phase behavior for $\kappa_{2.5}$ is representative of the generic phase diagram for an observable with a pair of tricritical points, such as $\kappa_2$. For $\kappa_{2.5}$ the region where $-\kappa_{2.5}$ differs from its lower convex envelope is $2/5<\rho_0<1$, and the inflection point is at $\rho_0\simeq0.6$.
	
	\begin{figure}[t]
		\includegraphics[scale=0.4]{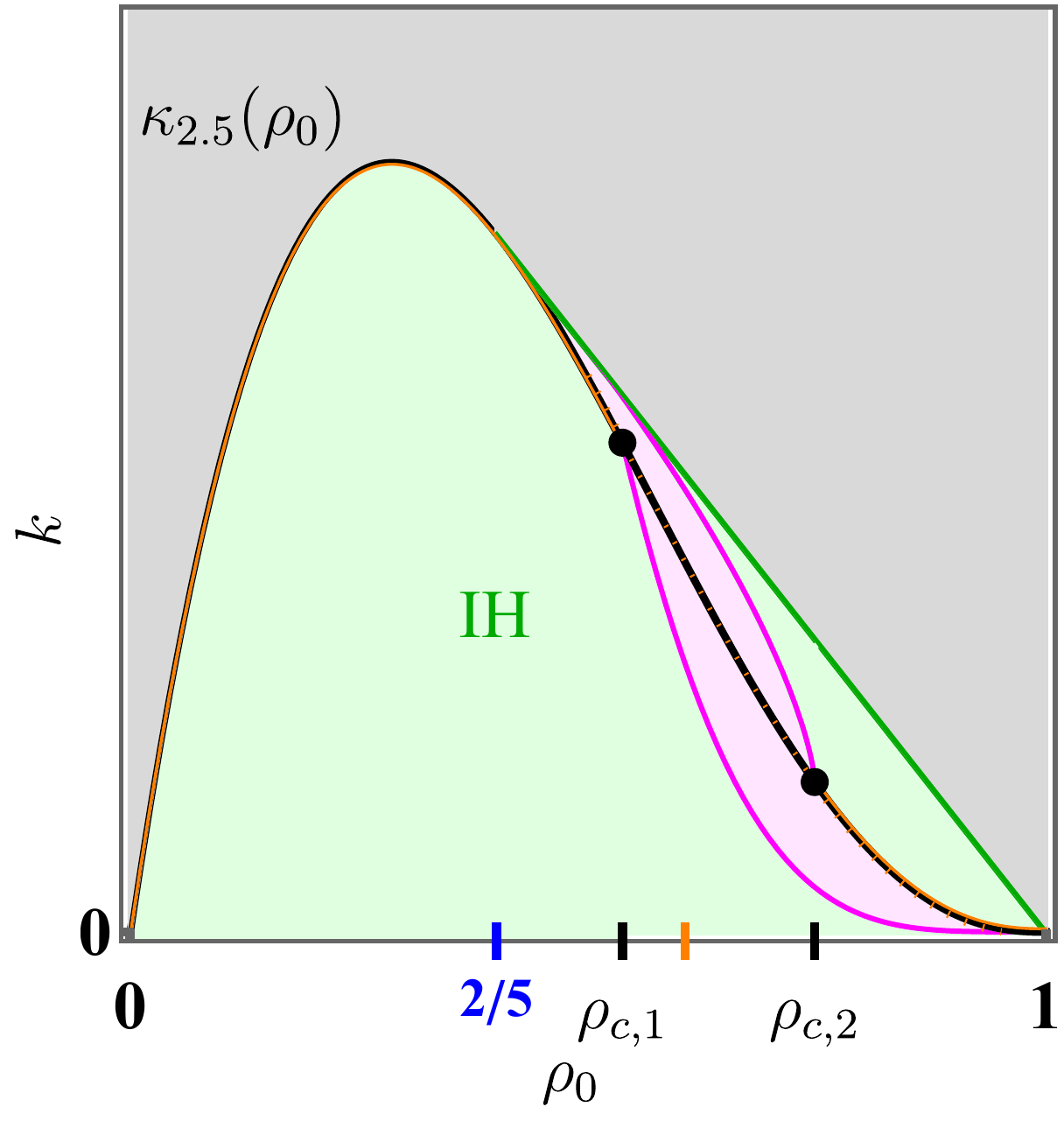}	
		\caption{Schematic phase diagram in the variables $(\rho_0,k)$ for the SSEP conditioned on $\kappa_{2.5}$. Miscibility gaps are denoted by  magenta shading. Black dots at $\rho_{c,1}$ and $\rho_{c,2}$ are tricritical points. They are placed on both sides of the inflection point $\kappa_{2.5}''=0$ (orange tick mark). The blue tick mark at $\rho_0=2/5$ indicate the boundaries of regions where $-\kappa_{2.5}$ differs from its lower convex envelope. Gray region is not accessible by hydrodynamic fluctuations.}
		\label{scheme}	
	\end{figure}
	
	We first discuss the parts of the $(\rho_0,k)$ phase diagram that are inaccessible via hydrodynamic mechanisms (the gray regions in Fig.~3, and Fig.~\ref{scheme}).
	From Eq.~(6), the accessible region is obtained by considering all possible $k$ and $\rho_0$ values that can be realized by a stationary profile $\rho(x)$
	\begin{equation}\label{krho}
	k=\int_0^1dx\kappa\left[\rho(x)\right] \quad,\quad\rho_0=\int_0^1dx\rho(x).
	\end{equation}
	These two relations define the convex hull of the curve $\kappa(\rho_0)$, as denoted by the green shading in Fig.~\ref{scheme} (or Fig.~3).  The density profiles $\rho(x)$ on the boundaries of this region are either homogeneous (with $k=\kappa(\rho_0)$ and rate function ${\cal I}=0$) or sharply phase-separated (with rate function ${\cal I}\to\infty$).
	Outside of this region the two constraints \eqref{krho} cannot be achieved for any density profile $\rho$.
	\footnote{For these values, the large deviation scaling behaviour is different, in fact
		$
		\log {\rm Prob}[K_T/(LT) \approx k] \simeq  -L^3T\, \mathcal I_{\rm m}(k)
		$ where $\mathcal I_{\rm m}$ is a different rate function.  It is important here that the trajectory duration $T$ is measured in hydrodynamic time units, so that $\hat{T} = L^2 T$ is the duration measured in microscopic units, so $L^3T\, \mathcal I_{\rm m}(k) = L \hat{T} \mathcal I_{\rm m}(k)$.  This is the scale for fluctuations that involve a change in the microscopic structure of the system, see for example~\cite{jack_hyperuniformity_2015}.}

	We now consider the phases and the miscibility gaps shown in Fig.~3 and Fig.~\ref{scheme}.  
	Recall that \eqref{kl} relates the $k$-values to corresponding values of the bias $\Lambda$.
	
	Note that for $k=\kappa(\rho_0)$, one may always solve $k=\Psi'(\Lambda)$ by taking $\Lambda=0$, which corresponds to a homogeneous (H) state.
	In fact, any homogeneous state that obeys \eqref{krho} must have exactly this value of $k$.  Hence the entire H phase in the biased ensemble (white regions in Fig.~1) must collapse to the line $k=\kappa(\rho_0)$ in the constrained ensemble.
	
	\begin{figure}[t]
		\includegraphics[scale=0.31]{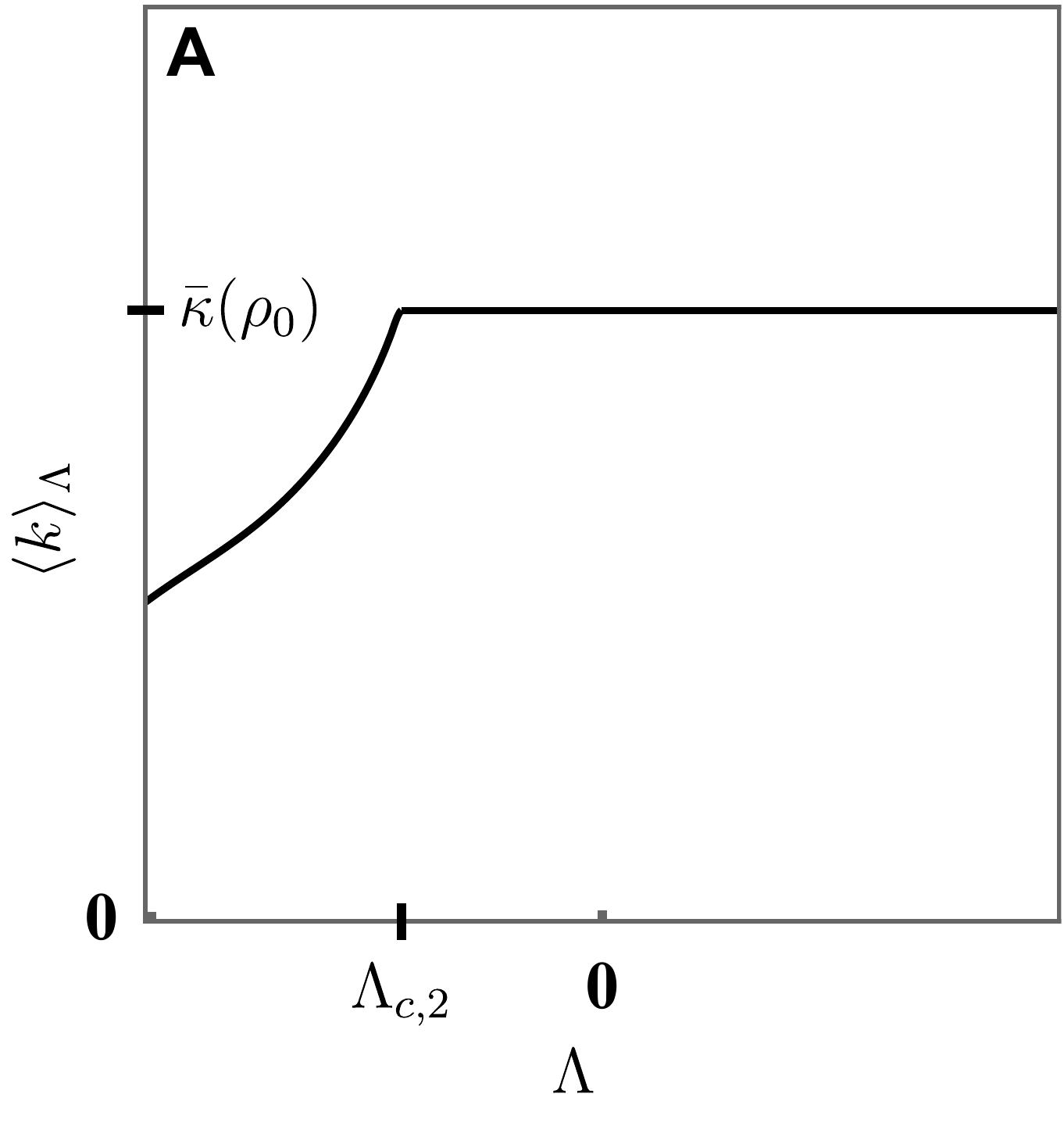}
		\includegraphics[scale=0.31]{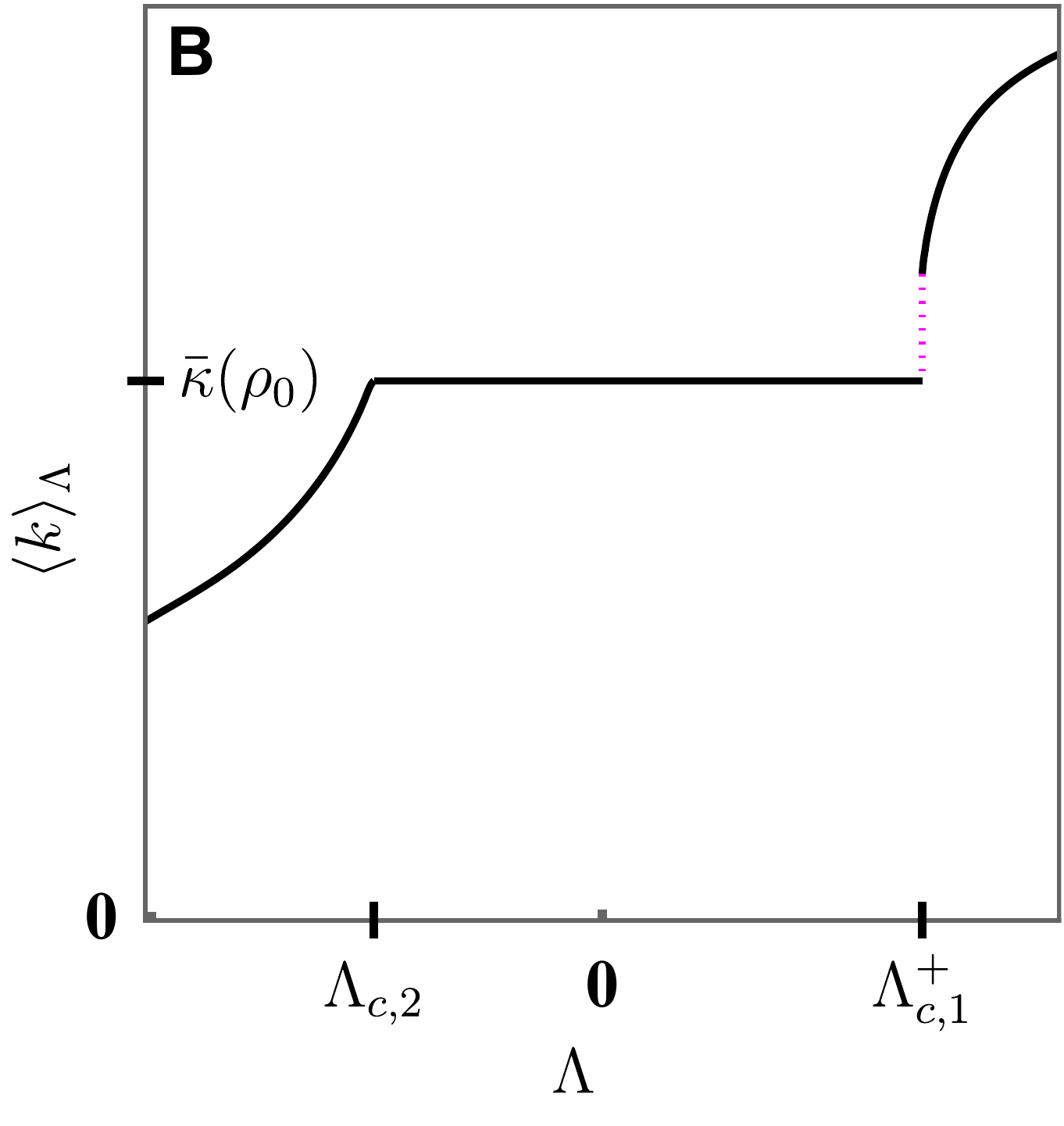}
		\includegraphics[scale=0.31]{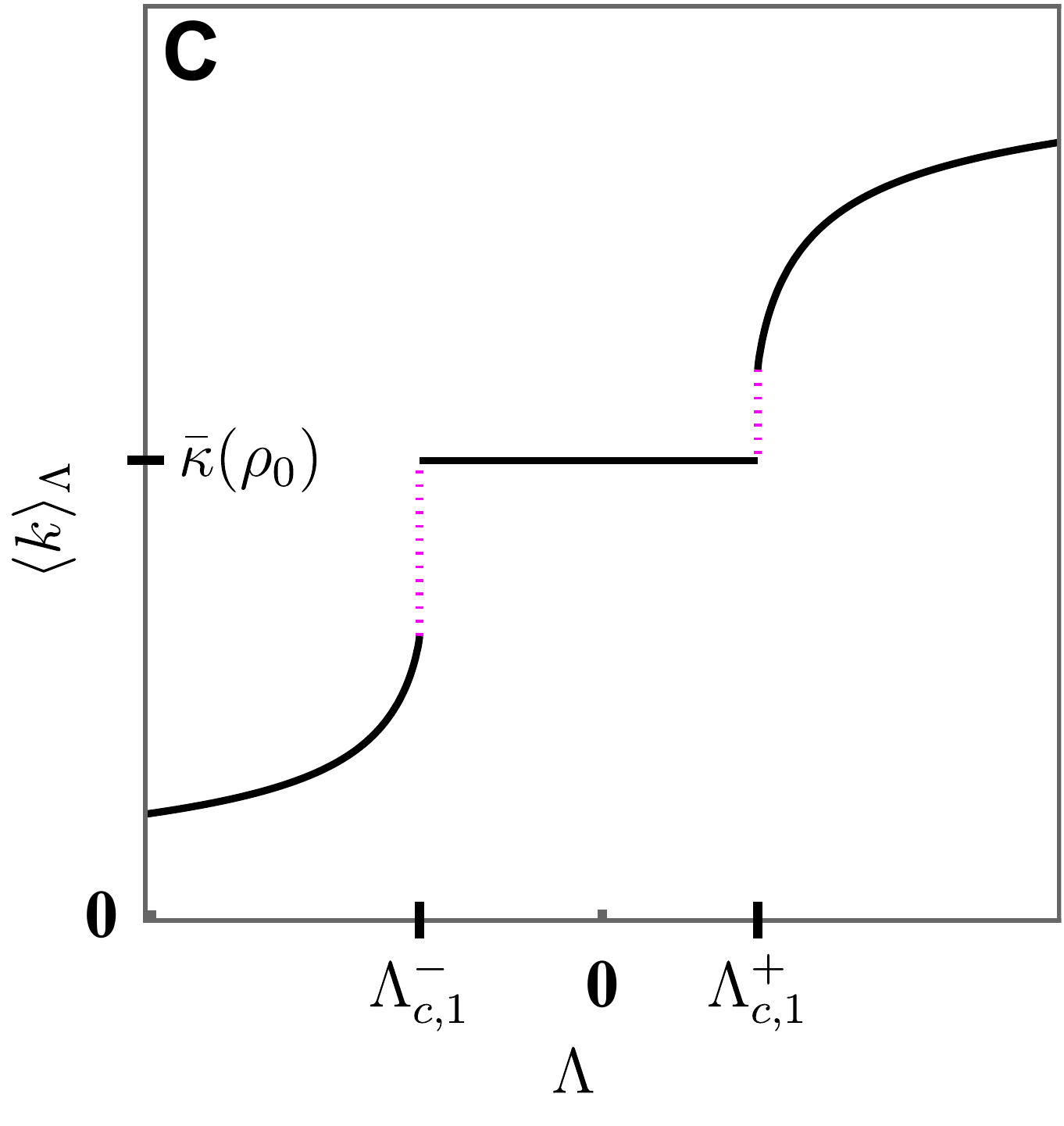}
		\includegraphics[scale=0.31]{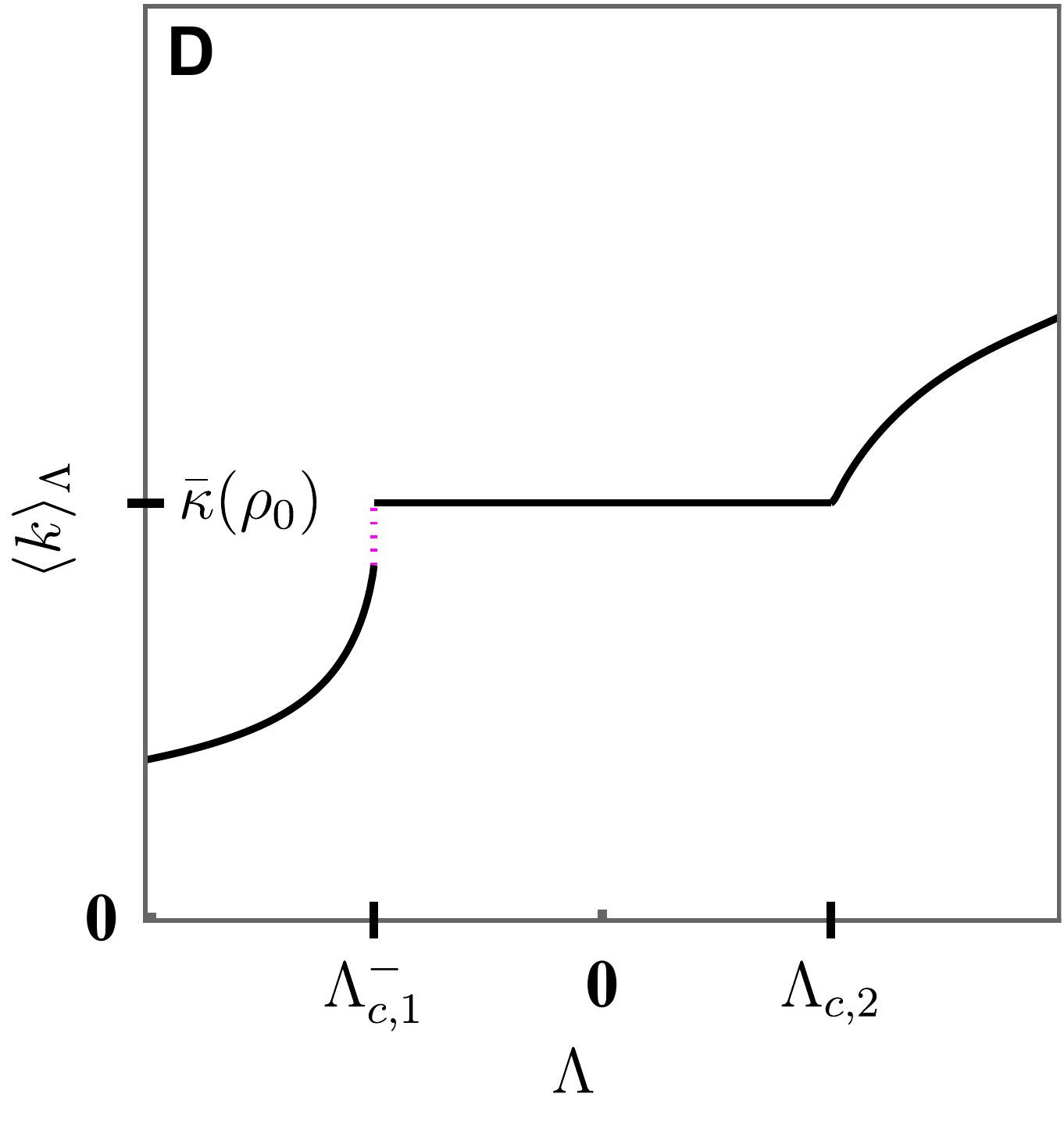}	
		\caption{The expectation $\langle k\rangle_{\Lambda}=\Psi'(\Lambda)$ \eqref{kl}, with respect to the biased ensemble \eqref{bias} at different values of the total density $\rho_0$: (A) $0<\rho_0<2/5$, (B) $2/5<\rho_0<\rho_{c,1}$, (C) $\rho_{c,1}<\rho_0<\rho_{c,2}$, (D) $\rho_{c,2}<\rho_0<1$. The  dotted lines indicate the miscibility gap which correspond to discontinuity in $\langle k\rangle_{\Lambda}=\Psi'(\Lambda)$. }
		\label{kappalambda}	
	\end{figure}
	
	For the transition into the IH phases and regimes of time-like phase separation there are different scenarios according to the value of $\rho_0$. These are determined by the the position of $\rho_0$ with respect to the tricritical points. As for $\kappa_2$, we find that for the SSEP biased by $\kappa_{2.5}$ there are two tricritical points at $\rho=\rho_{c,1},\rho_{c,2}$ (positioned on both sides of the inflection point), see Fig.~\ref{scheme}. As a result there are four different scenarios that we consider in the panels of Fig.~\ref{kappalambda}:
	\begin{enumerate}
		\item[A.] $0<\rho_0<2/5$\\ In this regime, $-\kappa_{2.5}$ is equal to its lower convex envelope, which means that $k>\kappa(\rho_0)$ is not hydrodynamically accessible (gray shading), and also $\Psi'(\Lambda)=\kappa(\rho_0)$ whenever $\Lambda>0$.  On the other hand, $\kappa_{2.5}$ differs from its lower convex envelope.  Also, $\Psi'$ is continuous, and deviates from $\kappa(\rho_0)$ for $\Lambda<\Lambda_{c,2}<0$.  By \eqref{kl}, this corresponds to a continuous H-IH transition in Fig.~\ref{scheme} (there is no time-like phase separation).  This is a ``single-sided'' transition since IH states only appear for $k-\kappa(\rho_0)<0$.
		\item[B.] $2/5<\rho_0<\rho_{c,1}$\\
		Both $\kappa_{2.5}$ and $-\kappa_{2.5}$ differ from their convex envelopes so DPTs must exist in the biased ensemble for both positive $\Lambda$ and negative $\Lambda$ (``double-sided'' transitions).
		We find $\kappa_{2.5}''<0$ so the transition for positive $\Lambda$ must be discontinuous.  By \eqref{kl}, this leads to time-like phase separation for $k>\kappa(\rho_0)$, shown in Fig.~\ref{scheme} by the pink miscibility gap.  On the other hand, the H-IH transition for $k<\kappa(\rho_0)$ is continuous in this range of density.
		\item[C.] $\rho_{c,1}<\rho_0<\rho_{c,2}$\\
		In this regime, both $\kappa_{2.5}$ and $-\kappa_{2.5}$ still differ from their convex envelopes so one still has double-sided behaviour. The resulting transitions are both discontinuous, so there are miscibility gaps on both sides of the line $k=\kappa(\rho_0)$ in Fig.~\ref{scheme}.
		\item[D.] $\rho_{c,2}<\rho_0<1$. \\
		The situation is similar to B, except that now the transition for positive $\Lambda$ is continuous and the one for negative $\Lambda$ is discontinuous.  Hence the miscibility gap in Fig.~\ref{scheme} lies below the line $k=\kappa(\rho_0)$.  
	\end{enumerate}

	\section{Relating the variational argument for H-IH transitions to the previous works~\cite{garrahan_dynamical_2007,garrahan_first-order_2009}}\label{H}	
	
	This Section points out a connection between the variational argument used here to establish discontinuous DPTs, and previous work in~ \cite{garrahan_dynamical_2007,garrahan_first-order_2009}.  It is not essential for the arguments of the main text, but it provides useful context.

	\subsection{Variational representation of the microscopic CGF}
	
	The authors of \cite{garrahan_dynamical_2007,garrahan_first-order_2009} exploited a variational formula for an CGF similar to $\Psi(\Lambda)$, to establish existence of discontinuous DPTs in kinetically constrained models.  We first define the variational formula, based on Donsker-Varadhan large deviation theory~\cite{donsker_asymptotic_1975-1,touchette_introduction_2018}.  The microscopic configuration of the model is denoted by $\eta=(\eta_1,\eta_2,\dots,\eta_L)$ where $\eta_i$ is the occupancy of the $i$th lattice site.  The transition rate from $\eta$ to $\eta'$ is $W(\eta',\eta)$ and we adopt the convention that $W(\eta,\eta) = -\sum_{\eta'}W(\eta',\eta)$.  Interpreting $W$ as a matrix, this means that its columns sum to zero. 
	To connect to the hydrodynamic arguments of the main text, we assume that the total particle number $N(\eta)=\sum_i \eta_i$ is conserved under the stochastic dynamics.
	
	Recall that the time variable $t$ used in this work is measured in hydrodynamic units. It is related to the microscopic time $\hat t$ as $t=\hat t/L^2$.  The arguments of \cite{garrahan_dynamical_2007,garrahan_first-order_2009} use microscopic units and we use $\hat{}$ to indicate this.  We consider trajectories of duration $\hat{T}$ (measured in microscopic units), and the analog of the observable $K_T$ is
	\begin{equation}
	K_{\hat{T}}\left(\cal X\right)=\int_0^{\hat{T}}d\hat{t} \,\hat{\kappa}(\mathbb\eta(\hat{t})).
	\end{equation}
	where $\hat\kappa$ is a suitable local observable. For example, consider the SSEP with $\hat{\kappa}=\hat{\kappa}_2$ with $\hat\kappa_2(\eta)=\sum_i \eta_i(1-\eta_{i+1})(1-\eta_{i-1})$. Then large deviations of this $K_{\hat{T}}$ correspond to large deviations of $K_T$ defined in Eq.~2 on taking $\kappa=\kappa_2$, at least for those fluctuations that take place by hydrodynamic mechanisms.

	Now consider the microscopic CGF
	\begin{equation}\label{psimic}
	\hat{\Psi}_L(s,\rho_0)=\frac{1}{L}\lim_{\hat{T}\to\infty} \frac{1}{\hat T} \log\langle e^{sK_{\hat{T}}}\rangle 
	\end{equation}
	where the average is taken in the steady state of the microscopic dynamics, at density $\rho_0$.  This object coincides~\cite{garrahan_dynamical_2007,garrahan_first-order_2009} with the largest eigenvalue of a matrix $W_\kappa$ whose elements are
	\begin{equation}
	W_{\hat{\kappa}}(\eta,\eta')=W(\eta,\eta')+s\hat{\kappa}(\eta)\delta_{\eta,\eta'} .
	\end{equation}  
	Note: since the total particle number is conserved, $W$ (and hence $W_{\hat{\kappa}}$) has a block-diagonal form where each block corresponds to a specific number of particles.  The CGF $\hat{\Psi}_L(s,\rho_0)$ is the largest eigenvalue of the block corresponding to the relevant number of particles $N=\rho_0 L$.

	To obtain a variational formula for this eigenvalue, we use that this matrix can be symmetrised.  For a generic model whose rates are in detailed balance with respect to an equilibrium probability distribution $P_{\rm eq}$, we write $\tilde{W}(\eta,\eta') = P_{\rm eq}^{-1/2}(\eta) W(\eta,\eta') P_{\rm eq}^{1/2}(\eta')$.   Detailed balance means that for $\eta\neq\eta'$ we have $\tilde{W}(\eta,\eta') = \sqrt{W(\eta,\eta') W(\eta',\eta)}$, so $\tilde{W}$ is symmetric.\footnote{
		We will take $P_{\rm eq}$ as a grand canonical equilibrium distribution so that $P_{\rm eq}(\eta)>0$ for all $\eta$.  The resulting $\tilde{W}$ does not depend on the value of the chemical potential of this distribution.}
	
	Then the Ritz variational formula for the largest eigenvalue of the relevant block of the matrix yields
	\begin{equation}\label{psi}
	\hat{\Psi}_L(s,\rho_0)=\frac{1}{L}\max_{V(\eta)}\frac{\sum_{\eta,\eta'} V(\eta)\left[\tilde{W}(\eta,\eta')+s\hat{\kappa}(\eta)\delta_{\eta,\eta'}\right]V(\eta')}{\sum_\eta V(\eta)^2}.
	\end{equation} 
	with the constraint that $V(\eta)=0$ if $N(\eta)\neq \rho_0 L$. At $s=0$ this maximum is achieved by the (canonical) equilibrium distribution $V(\eta)=\sqrt{P_{\text{eq}}(\eta)} \delta_{N(\eta),\rho_0L}$ which gives
	$\hat{\Psi}(0,\rho_0)=0$.  The insight of~\cite{garrahan_dynamical_2007,garrahan_first-order_2009} was that discontinuous DPTs can be established by considering the behaviour of \eqref{psi} for very small $s$.
	
	\subsection{Relating to the hydrodynamic limit}
	
	Comparing the definition of the CGF \eqref{scgf-def} with the microscopic CGF \eqref{psimic} and noting $\hat{T} = TL^2$ we have
	\begin{equation}\label{micmac}
	\Psi(\Lambda)=\lim_{L\to\infty}L^2\hat{\Psi}_L(\Lambda/L^2)
	\end{equation}
	That is, the bias parameter $s$ in the microscopic setting is related to the hydrodynamic bias $\Lambda$ as $s=\Lambda/L^2$, because of the
	hydrodynamic rescaling of time.  
	
	We will show that Eq.~(4) -- which is a variational representation of $\Psi$ -- is related to \eqref{psi}, which is a variational formula at microscopic level.  Using this relationship, we discuss conditions for existence of discontinuous DPTs.  We consider here the specific example of the SSEP, but the argument can be generalised quite easily.  As a suitable (grand-canonical) equilibrium distribution we take a product Bernoulli measure with mean density $\rho$, that is $P_{\rm eq}(\eta) = \prod_i \nu_\rho(\eta_i)$ where $\nu_\rho$ is the (marginal) distribution on each site.
	
	The analysis of lattice gas models in~\cite{garrahan_dynamical_2007,garrahan_first-order_2009} used a phase-separated state as variational ansatz in \eqref{psi}.  (This is phase separation in space, there should be no confusion with time-like phase separation.)  
	Write $\rho_l,\rho_h$ for the coexisting densities and 
	\begin{equation} \label{equ:y}
	y(\rho_l,\rho_h,\rho_0)=(\rho_h-\rho_0)/(\rho_h-\rho_l)
	\end{equation}
	for the fraction of the system that is occupied by the low density phase.  
	Then we take
	\begin{equation}\label{v}
	V(\eta)=\bigg[ \prod_{i=1}^{yL}\nu_{\rho_l}(n_i)\bigg]^{1/2} \bigg[\prod_{j=Ly+1}^{L}\nu_{\rho_h}(n_j) \bigg]^{1/2} \delta_{N(\eta),\rho_0 L}
	\end{equation}  
	as a variational ansatz corresponding to a phase-separated state with exactly $\rho_0 L$ particles.  
	
	\begin{figure}[t]
		\includegraphics[scale=0.4]{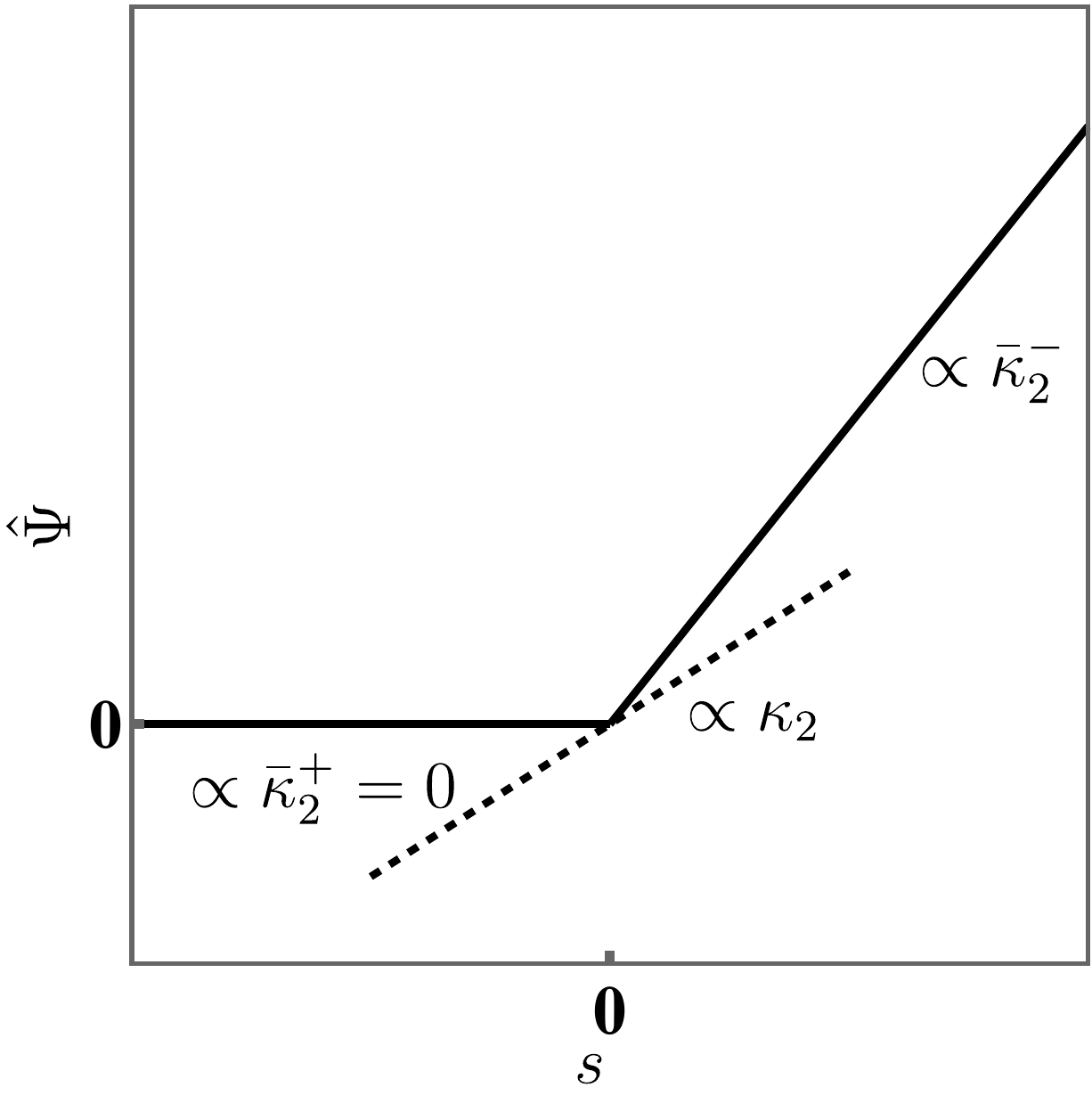}	
		\caption{A plot of the lower bound for $\hat{\Psi}(s)$, the right hand side of Eq.\eqref{lower2} for $\kappa_2$ at density $\rho_0=0.85$. Here $\kappa^+_2(\rho_0)=0$ is the lower convex envelope of $\kappa_2$ and $\kappa^-_2(\rho_0)=0.0375$ is the lower convex envelope of $-\kappa_2$ at $\rho_0$. The dashed line is the slope at the origin $\hat{\Psi}'(s=0)=\kappa_2(\rho_0)\simeq0.019$. }
		\label{psibound}	
	\end{figure}

	Plugging the test vector \eqref{v} into \eqref{psi}, it can easily be checked that the term involving $\tilde{W}$ yields a contribution of $O(1/L)$, because the system is locally equilibrated everywhere except in the vicinity of the two interfaces at $y=0$ and $i=yL$.  However, the term proportional to $s$ gives a contribution at $O(1)$, and the result is
	\begin{equation}\label{lower2}
	\hat{\Psi}_L(s,\rho_0)=\max_{\rho_l,\rho_h} \left\{ s  y(\rho_l,\rho_h,\rho_0) \kappa(\rho_l)+ s[1-y(\rho_l,\rho_h,\rho_0)] \kappa(\rho_h) \right\}  + O(1/L)
	\end{equation} 	
	where $\kappa(\rho)$ is the average of $\hat\kappa$ in an equilibrium state at density $\rho_0$.

	Maximising \eqref{lower2} over $\rho_l,\rho_h$ gives a convex envelope construction on $\kappa$ similar to the discussion in the main text.
	The result is that for $s\leq 0$, the maximum is achieved by coexistence between the densities that realize the lower convex envelope construction over $\kappa$; similarly for $s\geq 0$ one requires the lower convex envelope of $-\kappa$. However, exactly at $s=0$ the slope of $\hat{\psi}$ is given by the expected value \cite{touchette_large_2009}
	$
	\hat{\Psi}_L'(0,\rho_0)=\kappa(\rho_0)
	$.
	
	Denote by $\kappa^+(\rho_0)$ the lower convex envelope of $\kappa$, and similarly $\kappa^-(\rho_0)$ is the lower convex envelope of $-\kappa$.  
	The result is that (up to corrections at $O(1/L)$):
	\begin{align}
	\hat\Psi_L(s,\rho_0) & \geq s \kappa^-(\rho_0) , \qquad s>0
	\nonumber
	\\
	\hat\Psi_L(s,\rho_0) & \approx s \kappa(\rho_0) , \;\; \qquad |s| \lesssim L^{-2} 
	\\
	\nonumber
	\hat\Psi_L(s,\rho_0) & \geq s \kappa^+(\rho_0) , \qquad s<0
	\end{align}
	In cases where $\kappa$ (or $-\kappa$) differs from its convex envelope, this establishes discontinuities in $(\partial/\partial s) \lim_{L\to\infty} \Psi_L(s,\rho_0) $ at $s=0$.
	Such discontinuities were identified in~\cite{garrahan_dynamical_2007,garrahan_first-order_2009}  as DPTs. 
	An example is shown in Fig.~\ref{psibound} for $\kappa = \kappa_2$, where $\kappa$ differs from both $\kappa^\pm$.  
	
	The essential point is that the phase-separated density profiles that appear in this argument (and the corresponding convex envelopes) are exactly the same as those that appear in the discussion of the main text for the limits of $\Lambda\to\pm\infty$.  
	Recall that \eqref{micmac} indicates that $\Lambda=sL^2$. Then at  $L\to\infty$ one expects correspondence between the hydrodynamic behaviour at large $\Lambda$ and the microscopic behaviour at small non-zero $s$, consistent with the above analyses.

	\end{widetext}

\end{document}